\documentclass[prd,twocolumn,superscriptaddress]{revtex4-2}
\usepackage{amsmath,amssymb,amsfonts,amsthm}
\usepackage{graphics,graphicx,epsfig}
\usepackage{mathrsfs}
\usepackage{bm}
\usepackage{bbm}
\usepackage{upgreek}
\usepackage{longtable}
\usepackage{multirow}
\usepackage{array}
\usepackage{cases}
\usepackage{booktabs}
\usepackage[usenames,dvipsnames]{xcolor}

\usepackage{xcolor}
\usepackage[none]{hyphenat}
\usepackage{float}
\usepackage[a4paper,colorlinks=true,
linkcolor=blue,citecolor=blue,
pdfauthor={ },
pdftitle={ },
pdfsubject={ },
pdfkeywords={ }]{hyperref}

\newcommand{\note}[2]{}

\begin{document}

	\preprint{APS/123-QED}

	\title{Searching for Exotic Spin-Dependent Interactions with Diamond-Based Vector Magnetometer}

	\author{Aolong Guo}
	\affiliation{CAS Key Laboratory of Microscale Magnetic Resonance and School of Physical Sciences, University of Science and Technology of China, Hefei 230026, China}
	\affiliation{CAS Center for Excellence in Quantum Information and Quantum Physics, University of Science and Technology of China, Hefei 230026, China}

	\author{Runqi Kang}
	\affiliation{CAS Key Laboratory of Microscale Magnetic Resonance and School of Physical Sciences, University of Science and Technology of China, Hefei 230026, China}
	\affiliation{CAS Center for Excellence in Quantum Information and Quantum Physics, University of Science and Technology of China, Hefei 230026, China}
	\affiliation{Hefei National Laboratory, University of Science and Technology of China, Hefei 230088, China}

	\author{Man Jiao}
	\email{man.jiao@zju.edu.cn}
	\affiliation{Institute of Quantum Sensing and School of Physics, Zhejiang University, Hangzhou 310027, China}
	\affiliation{Institute for Advanced Study in Physics, Zhejiang University, Hangzhou 310027, China}

	\author{Xing Rong}
	\email{xrong@ustc.edu.cn}
	\affiliation{CAS Key Laboratory of Microscale Magnetic Resonance and School of Physical Sciences, University of Science and Technology of China, Hefei 230026, China}
	\affiliation{CAS Center for Excellence in Quantum Information and Quantum Physics, University of Science and Technology of China, Hefei 230026, China}
	\affiliation{Hefei National Laboratory, University of Science and Technology of China, Hefei 230088, China}

	\date{\today}

	\begin{abstract}

		We propose a new method to search for exotic spin-spin interactions between electrons and nucleons using a diamond-based vector magnetometer.
		The vector magnetometer can be constructed from ensembles of nitrogen-vacancy centers along different axes in a diamond.
		The $ ^{14}\mathrm{N} $ nuclear spins of nitrogen-vacancy centers in the same diamond can be polarized through the dynamic nuclear polarization method to serve as spin sources.
		With the vector magnetometer, the sought-after exotic interactions can be distinguished from the magnetic dipole-dipole interaction.
		For the axion-mediated interaction $ V_{PP} $, new upper bounds of the coupling $ |g_{P}^{e} g_{P}^{N}| $ are expected within the force range from 10 nm to 100 $ \upmu $m.
		For the $ Z' $-mediated interaction $ V_{AA} $, new upper bounds of the coupling $ |g_{A}^{e} g_{A}^{N}| $ are expected within the force range from 10 nm to 1 cm.
		The new upper bounds for $ V_{PP} $ and $ V_{AA} $ are both expected to be more than 5 orders of magnitude more stringent than existing constraints at the force range of 1 $ \upmu $m with the total measurement time of one day.

	\end{abstract}

	\maketitle

	\section{Introduction}

	Despite dark matter accounts for about 85\% of the matter in our Universe \cite{aghanim2020planck}, the nature of dark matter remains a mystery  \cite{particledatagroup2022review}.
	Axions are one of the well-motivated candidates of dark matter
	\cite{duffy2009axions,borsanyi2016calculation,dine2017axions,ballesteros2017unifying,chadha-day2022axion},
	and are also implied by the Peccei-Quinn solution of the strong $ CP $ problem \cite{peccei1977constraints,peccei1977mathrm,weinberg1978new,wilczek1978problem}.
	Exotic spin-dependent interactions between fermions can be mediated by axions \cite{moody1984new}.
	Subsequently, the mediating particles of the exotic interactions are extended to other new bosons beyond the Standard Model, such as $Z'$ bosons \cite{dobrescu2006spindependent,fadeev2019revisiting}.
	Therefore, searching for exotic interactions provides a promising approach for exploring new physics \cite{safronova2018search}.

	Experimental searches for exotic interactions have been performed using various precision measurement techniques \cite{safronova2018search},
	employing the torsion balance experiment \cite{terrano2015shortrange,hoedl2011improved,heckel2008preferredframe,ritter1990experimental}, trapped ions \cite{kotler2015constraints}, cantilever \cite{ding2020constraints}, spectroscopic measurements \cite{ficek2018constraints,ficek2017constraints}, atomic magnetometer \cite{wu2022experimental,almasi2020new,kim2019experimental,kim2018experimental,lee2018improved,ji2018new}, spin-based amplifier \cite{wang2022limits,su2021search}, and electric dipole moment measurements \cite{stadnik2018improved}.
	Single nitrogen-vacancy (NV) centers in diamond have been utilized as solid-state spin quantum sensors to search for exotic interactions at the micrometer scale \cite{maze2008nanoscale,degen2017quantum}.
	Ensembles of NV centers have also been utilized to search for exotic interactions as a high sensitivity magnetometer \cite{liang2023new,wu2023improved}.
	However, the explored force range is usually limited by the minimum distance from each spin sensor to the nearest spin source,
	because exotic interactions decay exponentially with increasing distance.
	In addition, searching for the static exotic spin-spin interactions is a challenge, due to the difficulty of eliminating spurious effects of the accompanying magnetic dipole-dipole interaction.

	Here we propose searching for the static exotic spin-spin interactions $ V_{PP} $ and $ V_{AA} $ between electrons and nucleons with a diamond-based vector magnetometer.
	NV centers in diamond are aligned along the four crystallographic axes.
	The electron spins of NV centers along three different axes can be utilized to construct the vector magnetometer.
	The $ ^{14}\mathrm{N} $ nuclear spins of NV centers along the fourth axis can be polarized through the dynamic nuclear polarization (DNP) method to serve as the spin sources.
	The vector magnetometer can distinguish the possible effective magnetic field induced by the exotic interaction and the dipole magnetic field by their different directions.
	With such configurations, the nanoscale distance from each spin sensor to the nearest spin source allows for exploring the force range at the nanometer scale.

	The structure of this paper is as follows.
	In Sec.~\ref{sec:Scheme}, the experimental scheme of the proposal is demonstrated.
	In Sec.~\ref{sec:Analysis_and_Results}, the expected signal of the exotic interactions and the magnetic dipole-dipole interaction are calculated, and the expected constraints of the exotic interactions are presented.
	The conclusion is given in Sec.~\ref{sec:Conclusion}.

	\section{Experimental Scheme \label{sec:Scheme}}

	\subsection{Effective Magnetic Fields from Exotic Spin-Spin Interactions}

\begin{figure}
		\includegraphics[width=\linewidth]{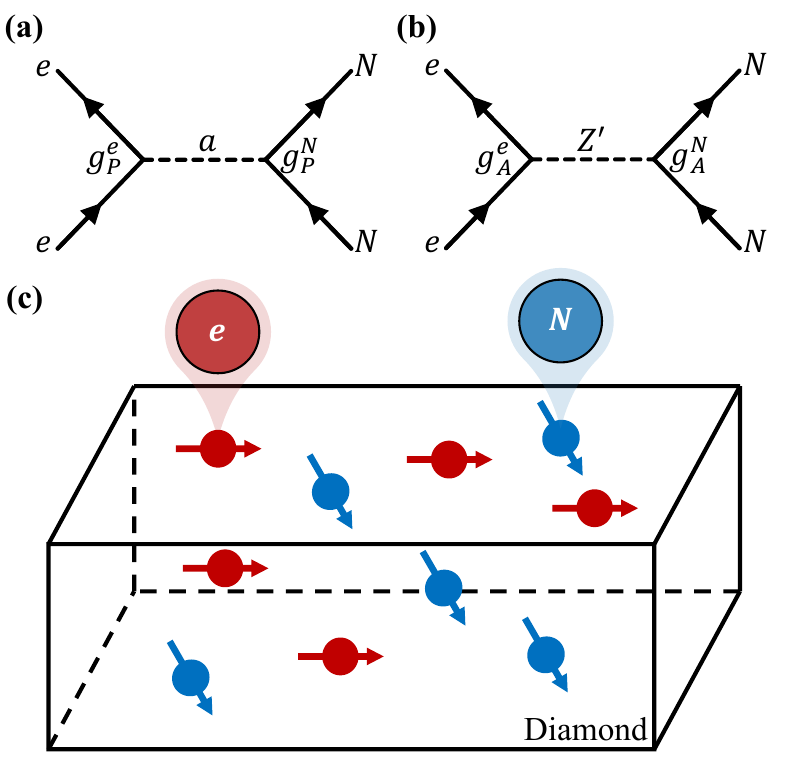}
		\caption{\label{fig:Feynman}
			Schematic of the experimental proposal.
			(a) Feynman diagram of the axion-mediated exotic interaction $ V_{PP} $ between an electron and a nucleon.
			(b) Feynman diagram of the $ Z' $-mediated exotic interaction $ V_{AA} $ between an electron and a nucleon.
			(c) Scheme of the proposed experimental search for the exotic interactions.
			The nucleon spins of NV centers along one axis can be used as the spin sources.
			The electron spins of NV centers along other axes can be used as the spin sensors.
		}
	\end{figure}

	The non-relativistic potentials of the targeted exotic interactions are \cite{dobrescu2006spindependent}
	\begin{subequations}
		\begin{align}
			V_{PP} =
			& - \dfrac{g_{P}^{e} g_{P}^{N} \hbar^{3}}{16 \pi  m_{e} m_{N} c}
			\left [
			( \bm{\sigma}_{e} \cdot \bm{\sigma}_{N} )
			( \dfrac{1}{r^{3}} + \dfrac{1}{\lambda r^{2}} )
			\right.
			\notag\\
			& - \left .
			( \bm{\sigma}_{e} \cdot \bm{e}_{r} )
			( \bm{e}_{r} \cdot \bm{\sigma}_{N} )
			( \dfrac{3}{r^{3}} + \dfrac{3}{\lambda r^{2}} + \dfrac{1}{\lambda^{2} r} )
			\right ]
			e^{-r/\lambda},\label{eqs:VPP}
			\\
			V_{AA} =
			& - \dfrac{g_{A}^{e} g_{A}^{N} \hbar c}{4 \pi}
			( \bm{\sigma}_{e} \cdot \bm{\sigma}_{N} )
			\dfrac{e^{-r/\lambda}}{r},\label{eqs:VAA}
		\end{align}
	\end{subequations}
	where
	$ \bm{\sigma}_{e} $ ($ \bm{\sigma}_{N} $) is the unit vector of the interacting electron (nucleon) spin,
	$ \lambda = \hbar / m_{X} c $ is the force range with the mass $ m_{X} $ of the mediating boson ($ X $ refers to axions and $Z'$ bosons for $ V_{PP} $ and $ V_{AA} $, respectively),
	$ r $ is the distance between the interacting spins,
	$ \bm{e}_{r} $ is the direction from the nucleon spin to the electron spin,
	$ m_{e} $ is the electron mass and
	$ m_{N} $ is the nucleon mass.
	Here,
	$ g_{P}^{e} $ and $ g_{P}^{N} $ are the dimensionless pseudo-scalar couplings of axions to electrons and nucleons, respectively \cite{moody1984new,dobrescu2006spindependent,fadeev2019revisiting},
	while $ g_{A}^{e} $ and $ g_{A}^{N} $ are the dimensionless axial-vector couplings of $Z'$ bosons to electrons and nucleons, respectively \cite{fayet1986fifth,dobrescu2006spindependent,fadeev2019revisiting}.
	The couplings to nucleons are assumed to be the same for neutrons and protons.
	The Feynman diagrams are shown in Fig.~\ref{fig:Feynman} (a), (b).

	The induced effective magnetic fields on the electron spin are
	\begin{subequations}
		\begin{align}
			\bm{B}_{\mathrm{eff},PP} =
			& - \dfrac{g_{P}^{e} g_{P}^{N} \hbar^{2}}{8\pi m_{e} m_{N} c \gamma_{e}}
			\left [
			\bm{\sigma}_{N}
			( \dfrac{1}{r^{3}} + \dfrac{1}{\lambda r^{2}} )
			\right.
			\notag\\
			& - \left .
			\bm{e}_{r}
			( \bm{e}_{r} \cdot \bm{\sigma}_{N} )
			( \dfrac{3}{r^{3}} + \dfrac{3}{\lambda r^{2}} + \dfrac{1}{\lambda^{2} r} )
			\right ]
			e^{-r/\lambda},\label{eqs:BeffPP}
			\\
			\bm{B}_{\mathrm{eff},AA} =
			& - \dfrac{g_{A}^{e} g_{A}^{N} c}{2 \pi \gamma_{e}} \bm{\sigma}_{N} \dfrac{e^{-r/\lambda}}{r},\label{eqs:BeffAA}
		\end{align}
	\end{subequations}
	where $ \gamma_{e} $ is the electron gyromagnetic ratio.

	We propose utilizing electron spins and $ ^{14}\mathrm{N} $ nuclear spins of ensembles of NV centers inside one diamond as the spin sensors and the spin sources, respectively, as shown in Fig.~\ref{fig:Feynman} (c).
	The averaged effective magnetic fields $ \bm{b}_{\mathrm{eff},PP} $ and $ \bm{b}_{\mathrm{eff},AA} $ on the ensembles of spin sensors
	can be derived from integrating $ \bm{B}_{\mathrm{eff},PP} $ and $ \bm{B}_{\mathrm{eff},AA} $, respectively, over both the electron spins and the nuclear spins.

	In this proposed search for the exotic interactions, the magnetic dipole-dipole interaction is the main spurious effect.
	The dipole magnetic field on the electron spin is
	\begin{equation}\label{eqs:Bdip}
		\bm{B}_{\mathrm{dip}} = \dfrac{\mu_{0}}{4\pi} \gamma_{n} \hbar [3\bm{e}_{r}(\bm{e}_{r}\cdot\bm{\sigma}_{n}) - \bm{\sigma}_{n}] \dfrac{1}{r^{3}},
	\end{equation}
	where
	$ \bm{\sigma}_{n} $ is the unit vector of the nuclear spin,
	$ \mu_{0} $ is the vacuum permeability and
	$ \gamma_{n} $ is the $ ^{14}\mathrm{N} $ nuclear gyromagnetic ratio.
	The Fermi contact interaction term is neglected, since NV centers are spatially separated.
	The averaged dipole magnetic field $ \bm{b}_{\mathrm{dip}} $ on the ensembles of spin sensors can be derived from integrating $ \bm{B}_{\mathrm{dip}} $ over both the electron spins and the nuclear spins.

	\subsection{Diamond-Based Vector Magnetometer}

	\begin{figure*}
		\includegraphics[width=\linewidth]{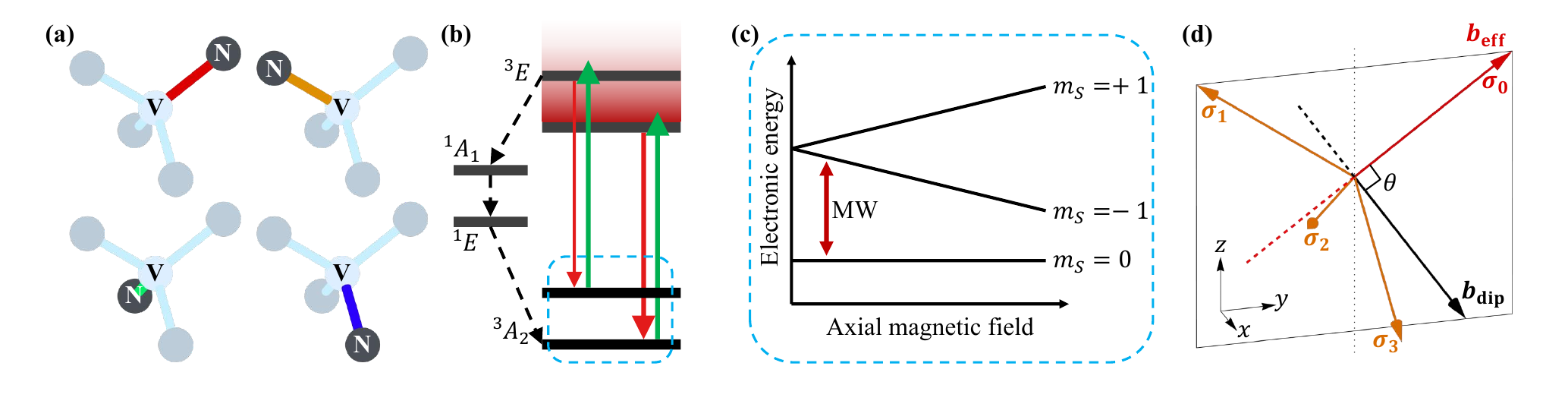}
		\caption{\label{fig:NV}
			Diamond-based vector magnetometer.
			(a) NV centers aligned with the four crystallographic axes in a single-crystal diamond.
			(b) Energy-level diagram of an NV center in diamond \cite{schirhagl2014nitrogenvacancy}.
			The NV center can be pumped by green laser from the ground state $ ^{3} A_{2} $ to the excited state $ ^{3} E $.
			The decay back to $ ^{3} A_{2} $ via singlet states $ ^{1} A_{1} $ and $ ^{1} E $ transfers the populations from $ | m_{S} = \pm 1 \rangle $ to $ | m_{S} = 0 \rangle $.
			(c) The Zeeman shift of the spin sublevels of the ground state $ ^{3} A_{2} $.
			The transition between $ | m_{S} = 0 \rangle $ and $ | m_{S} = -1 \rangle $ can be driven by a resonant microwave.
			(d) The principle of distinguishing the possible effective magnetic field and the dipole magnetic field using the vector magnetometer.
			The possible effective magnetic field induced by the exotic interaction $V_{PP}$ or $V_{AA}$ is denoted by $ \bm{b}_{\mathrm{eff}} $,
			while the dipole magnetic field induced by the magnetic dipole-dipole interaction is denoted by $ \bm{b}_{\mathrm{dip}} $.
			The angle between the fields is $ \theta $.
			For the magnetic-field vector that can be obtained from the vector magnetometer, the projection on the red dashed line only contains the information of $ \bm{b}_{\mathrm{eff}} $,
			while the projection on the black dashed line only contains the information of $ \bm{b}_{\mathrm{dip}} $.
		}
	\end{figure*}

	NV centers are paramagnetic point defects in diamond.
	An NV center consists of a substitutional nitrogen atom and a neighbouring vacancy \cite{gruber1997scanning,jelezko2006single}.
	Since there are four crystallographic axes of C-C bonds in the diamond lattice, NV centers in diamond can be classified into four groups, as shown in Fig.~\ref{fig:NV}(a).
	The axes of NV centers are denoted by $\bm{\sigma}_{0}, \bm{\sigma}_{1}, \bm{\sigma}_{2}$, and $\bm{\sigma}_{3}$.

	Figure \ref{fig:NV}(b) shows the electronic energy levels of an NV center \cite{doherty2012theorya,gali2019initio,doherty2013nitrogenvacancy}.
	The spin state can be optically initialized into $ | m_{S} = 0 \rangle $ and read out by green laser \cite{doherty2013nitrogenvacancy,goldman2015phononinduced,goldman2015stateselective}.
	The resonant frequency of manipulation microwave is sensitive to the magnetic-field projection on the NV axis, as shown in Fig.~\ref{fig:NV}(c) \cite{schirhagl2014nitrogenvacancy,degen2017quantum}.

	By applying a carefully designed external magnetic field,
	NV centers with different axes can be manipulated and read out separately.
	The NV centers along $\bm{\sigma}_{1}, \bm{\sigma}_{2}, \bm{\sigma}_{3}$ can be used to construct the diamond-based vector magnetometer \cite{maertz2010vector,pham2011magnetica,schloss2018simultaneous}.
	The $ ^{14}\mathrm{N} $ nuclear spins of NV centers along the axis $ \bm{\sigma}_{0} $ can be polarized, in order to generate the possible effective magnetic field $ \bm{b}_{\mathrm{eff}} $ induced by the exotic interaction $V_{PP}$ or $V_{AA}$ \cite{smeltzer2009robust,pagliero2014recursive,soshenko2021nuclear}.
	According to the simulations discussed in Sec.~\ref{subsec:Vexo} and Sec.~\ref{subsec:Vdip}, the accompanying dipole magnetic field $ \bm{b}_{\mathrm{dip}} $ is expected to be perpendicular to $ \bm{b}_{\mathrm{eff}} $ with the considered parameters, as shown in Fig.~\ref{fig:NV}(d).
	Since the vectors $\bm{\sigma}_{1}, \bm{\sigma}_{2}, \bm{\sigma}_{3}$ span the 3-dimensional space,
	we can obtain the projection of the measured magnetic field on any arbitrary direction by combining the signals from the three groups of NV centers.
	Therefore, the possible effective magnetic field $ \bm{b}_{\mathrm{eff}} $ and the dipole magnetic field $ \bm{b}_{\mathrm{dip}} $ can be distinguished by the vector magnetometer.

	\subsection{Pulse Sequence}

\begin{figure}
		\includegraphics[width=\linewidth]{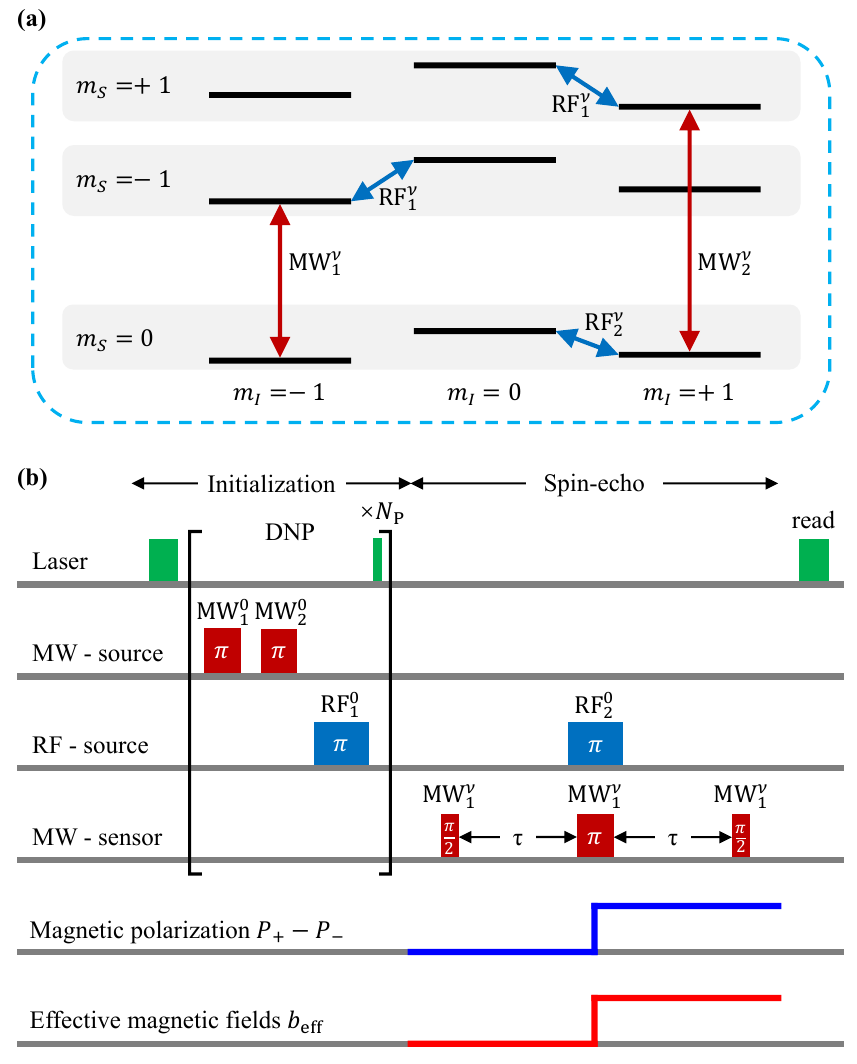}
		\caption{\label{fig:sequence}
			(a) Detailed energy levels of the ground state of a $ ^{14} $NV center.
			The red (blue) arrows denote the microwave (radio-frequency) pulses applied in the sequence.
			(b) Pulse sequence of the proposed experiment.
			The first laser pulse can initialize the electron spin of NV centers along all axes to the state $ | m_{S} = 0 \rangle $.
			The DNP sequence can recursively initialize the $ ^{14}\mathrm{N} $ nuclear spins of NV centers along axis $ \bm{\sigma}_{0} $ to the state $ | m_{I} = 0 \rangle $.
			The population can be transferred from $|m_{I} = 0\rangle$ to $|m_{I} = 1\rangle$ by the RF$^{0}_{2}$ $\pi$ pulse, in order to obtain the magnetic polarization and to generate a possible effective magnetic field.
			The following spin-echo sequence is used to measure the possible effective magnetic field.
			The sequence is repeated sequentially with $\nu = 1, 2, 3$ to perform a time-division multiplexing vector measurement.
		}
	\end{figure}

	Microwave (MW) and radio-frequency (RF) pulses are used to manipulate the electron spins and the nuclear spins of NV centers, respectively.
	Figure \ref{fig:sequence}(a) shows the transitions engaged in the pulse sequence.
	The subscripts 1 and 2 of MWs and RFs indicate different transitions, while the superscript $\nu = 0,1,2,3$ indicates the NV axis.

	The pulse sequence of the proposed experiment is shown in Fig.~\ref{fig:sequence}(b).
	A laser pulse can initialize electron spins of NV centers along all axes into $ | m_{S} = 0 \rangle $.
	After the electron spin initialization, a DNP sequence can be used to polarize the $ ^{14}\mathrm{N} $ nuclear spins in NV centers along $ \bm{\sigma}_{0} $ \cite{smeltzer2009robust,pagliero2014recursive,soshenko2021nuclear}.
	The populations of the states $ | m_{S} = 0, m_{I} = \pm 1 \rangle $ are firstly transferred to the states $|m_{S} = \pm 1, m_{I} = \pm 1\rangle$ by spectrally narrow $\pi$ pulses of MW$^{0}_{1}$ and MW$^{0}_{2}$,
	and then transferred to the states $ | m_{S} = \pm 1, m_{I} = 0 \rangle $ by a spectrally broad RF$^{0}_{1}$ $\pi$ pulse.
	Finally, a short laser pulse repolarizes the electron spin.
	Such a DNP sequence can be repeated to recursively obtain an $ 80\% $ or even higher population $ P_{0} $ of the state $ | m_{I} = 0 \rangle $ \cite{pagliero2014recursive,soshenko2021nuclear,soshenko2023optimal}.

	A following spin-echo sequence \cite{hahn1950spin} can be used to obtain a sensitivity to the possible effective magnetic field $ \bm{b}_{\mathrm{eff}} $.
	The MW$^{\nu}_{1}$ pulses here are resonant with NV centers along $ \bm{\sigma}_{\nu} $, where $ \nu = 1,2,3 $ denotes the NV axis of the sensors.
	At the same time as the MW$^{\nu}_{1}$ $\pi$ pulse, an RF$^{0}_{2}$ $\pi$ pulse for NV centers along $ \bm{\sigma}_{0} $ can transfer the population from $|m_{I} = 0\rangle$ to $|m_{I} = 1\rangle$.
	After such a pulse, the magnetic polarization $ \langle m_{I} \rangle = P_{+} - P_{-} $ of the $ ^{14}\mathrm{N} $ nuclear spins can reach $ 70\% $ or even higher, where $ P_{\pm} $ are the populations of $ | m_{I} = \pm 1 \rangle $.
	The magnetic polarization can generate the possible effective magnetic field on the sensors.
	As a result, the axial projection of the possible effective magnetic field can be encoded into the population of the state $|m_{S} = 0\rangle$ of the sensors at the end of the spin-echo sequence
	and can be read out optically.
	On the contrary, the influences due to low-frequency environmental noise can be eliminated.
	The duration of the spin-echo sequence should be much longer than the RF$^{0}_{2}$ $\pi$ pulse but shorter than the coherence time $ T_{2} $ of the NV ensemble.
	Therefore, the concentration of NV centers in the diamond is limited by the requirement of a sufficiently long coherence time \cite{bauch2020decoherence,schloss2018simultaneous}.

	In order to obtain all the magnetic signals from NV centers along the axes $\bm{\sigma}_{1}, \bm{\sigma}_{2}, \bm{\sigma}_{3}$,
	the pulse sequence shown in Fig.~\ref{fig:sequence} can be repeated sequentially with $\nu = 1, 2, 3$.

	\section{Analysis and Results \label{sec:Analysis_and_Results}}

	\subsection{Simulation of the Exotic Interactions \label{subsec:Vexo}}

	The Monte-Carlo method is used to calculate the averaged effective magnetic fields $ \bm{b}_{\mathrm{eff},PP} $ and $ \bm{b}_{\mathrm{eff},AA} $ on the ensembles of spin sensors.
	In such a calculation, the spin sources are supposed to be aligned to the axis $ \bm{\sigma}_{0} $.
	The lower cutoff of the considered force range is 1 nm, which should be much longer than the typical 0.15-nm C-C bond length in diamond.
	The upper cutoff of the considered force range of $ \bm{b}_{\mathrm{eff},PP} $ should be much shorter than the scale of the diamond.
	The integration of Eqs.~\ref{eqs:BeffPP} can be simplified with such force ranges, since the former four of the five terms cancel out after the integration.
	Therefore, the size of the diamond is considered to be $ 10 \times 10 \times 2 \ \mathrm{mm}^{3} $, and the corresponding upper cutoff of force range of $ \bm{b}_{\mathrm{eff},PP} $ is 100 $ \upmu $m.
	For a typical force range of $\lambda = 10$ nm, the results are
	\begin{subequations}
		\begin{align}
			\bm{b}_{\mathrm{eff},PP}^{\mathrm{sim}} = 3.5 \ \mathrm{fT} \times
			(\frac{g_{P}^{e}g_{P}^{N}}{10^{-4}}) (\frac{P_{+} - P_{-}}{70\%}) (\frac{\zeta}{0.5})
			\bm{\sigma}_{0}, \label{eqs:sim,exoPP}
			\\
			\bm{b}_{\mathrm{eff},AA}^{\mathrm{sim}} = 5.2 \ \mathrm{fT} \times
			(\frac{g_{A}^{e}g_{A}^{N}}{10^{-17}}) (\frac{P_{+} - P_{-}}{70\%}) (\frac{\zeta}{0.5})
			\bm{\sigma}_{0}. \label{eqs:sim,exoAA}
		\end{align}
	\end{subequations}

	It is noteworthy that the strengths of the effective magnetic fields should be multiplied by a factor $\zeta = 0.5$ \cite{engel1989spindependent,kimball2015nucleara},
	which is the proportion of the $ ^{14}\mathrm{N} $ nuclear spin angular momentum contributed from the proton spins and the neutron spins,
	since the orbital angular momenta of the protons and the neutrons are not engaged in the exotic interactions.
	Both of the effective magnetic fields $ \bm{b}_{\mathrm{eff},PP} $ and $ \bm{b}_{\mathrm{eff},AA} $ are in the direction of $\bm{\sigma}_{0} = (0,\sqrt{2/3},\sqrt{1/3})$, parallel to the axis of spin sources.

	\subsection{Simulation of the Magnetic Dipole-Dipole Interaction \label{subsec:Vdip}}

	The finite element method is used to numerically simulate the averaged dipole magnetic field $ \bm{b}_{\mathrm{dip}} $ on the ensembles of spin sensors.
	With the same parameters as the calculation of the exotic interactions, the finite-element-method simulation gives
	\begin{equation}\label{eqs:bPP}
		\bm{b}_{\mathrm{dip}}^{\mathrm{sim}} = 2.0 \ \mathrm{pT} \times
		(\frac{\gamma_{n}}{2\pi\times3.077 \ \mathrm{MHz/T}}) (\frac{P_{+} - P_{-}}{70\%})
		\bm{e}_{n},
	\end{equation}
	where $ \bm{e}_{n} = (0,\sqrt{1/3},-\sqrt{2/3}) $ is the direction of the dipole magnetic field, as shown in Fig.~\ref{fig:NV}(d).
	The numerical error of $ \bm{b}_{\mathrm{dip}}^{\mathrm{sim}} \cdot \bm{\sigma}_{0} $ is estimated to be no more than 2 fT, which is much smaller than the measurement uncertainty.

	\subsection{Expected Constraints}

	In the proposal, the axial magnetic fields experimentally measured by the three groups of sensors of the vector magnetometer are denoted by $ b_{1}, b_{2} $ and $ b_{3} $.
	Here we take the exotic interaction $V_{PP}$ for example.
	By linearly combining these measured results properly, we can separately derive the magnitude of $ \bm{b}_{\mathrm{eff},PP} $ and the magnitude of $ \bm{b}_{\mathrm{dip}} $:
	\begin{subequations}
		\begin{align}
			b_{\mathrm{eff},PP}^{\mathrm{exp}} &
			= \dfrac{\xi_{1}b_{1} + \xi_{2}b_{2} + \xi_{3}b_{3}}{\sin\theta},\label{exp,exo}
			\\
			b_{\mathrm{dip}}^{\mathrm{exp}} &
			= \dfrac{\zeta_{1}b_{1} + \zeta_{2}b_{2} + \zeta_{3}b_{3}}{\sin\theta},\label{exp,dip}
		\end{align}
	\end{subequations}
	where
	$ \xi_{1} = -1 $, $ \xi_{2} = -1 $ and $ \xi_{3} = -1 $ are the combination coefficients for $ b_{\mathrm{eff},PP} $,
	$ \zeta_{1} = -0.707 $, $ \zeta_{2} = 0.354 $ and $ \zeta_{3} = 0.354 $ are the combination coefficients for $ b_{\mathrm{dip}} $,
	and $ \theta = 90^{\circ} $ is the angle between $\bm{b}_{\mathrm{eff},PP}$ and $\bm{b}_{\mathrm{dip}}$.

	The estimated uncertainties of the derived fields are
	\begin{subequations}
		\begin{align}
			\delta b_{\mathrm{eff},PP}^{\mathrm{exp}} = \dfrac{\sqrt{\xi_{1}^{2}+\xi_{2}^{2}+\xi_{3}^{2}}}{\sin\theta} \dfrac{\eta}{\sqrt{T}},\label{delta,exp,exo}\\
			\delta b_{\mathrm{dip}}^{\mathrm{exp}} = \dfrac{\sqrt{\zeta_{1}^{2}+\zeta_{2}^{2}+\zeta_{3}^{2}}}{\sin\theta} \dfrac{\eta}{\sqrt{T}},\label{delta,exp,dip}
		\end{align}
	\end{subequations}
	where the sensitivity $ \eta = 1.3 \ \mathrm{pT/\sqrt{Hz}} $ to $ b_{1}, b_{2} $ and $ b_{3} $ is estimated by shot noise limit \cite{barry2020sensitivity}:
	\begin{equation}\label{key}
		\eta =
		\dfrac{\pi}{2}
		\dfrac{1}{\gamma_{e}}
		\dfrac{1}{\sqrt{N \tau_{a}}}
		\dfrac{1}{e^{-(\tau_{a} / T_{2})^{p}}}
		\sigma_{R}
		\sqrt{\dfrac{t_{\mathrm{tot}}}{\tau_{a}}},
	\end{equation}
	where
	$ \gamma_{e} $ is the electron gyromagnetic ratio,
	$ N = 8.8 \times 10^{14} $ is the number of NV centers along a single axis in the ensemble,
	the phase accumulation time $ \tau_{a} = 2 \tau $ is equal to the decoherence time $ T_{2} \approx 160 \ \mathrm{\upmu s} $ \cite{bauch2020decoherence},
	$ p $ is a stretched exponential parameter ranging from 0.5 to 3 \cite{barry2020sensitivity},
	$ \sigma_{R} \sim 5000 $ is the spin-readout noise \cite{schloss2018simultaneous},
	$ t_{\mathrm{tot}} = 3 (t_{I} + \tau_{a} + t_{R}) $ is the total time of one cycle of three-axes measurements,
	$ t_{I} \approx 657 \ \mathrm{\upmu s} $ is the time of the initialization pulse sequence in Fig.~\ref{fig:sequence} \cite{soshenko2021nuclear},
	and $ t_{R} \approx 10 \ \mathrm{\upmu s} $ is the readout time \cite{wolf2015subpicotesla}.
	Given the total measurement time $ T = 1 \ \mathrm{day} $, the uncertainty $ b_{\mathrm{eff},PP}^{\mathrm{exp}} = 7.7 \ \mathrm{fT} $ is expected.
	Besides, an experimentally achievable sensitivity $ \eta = 50 \ \mathrm{pT/\sqrt{Hz}} $ and the corresponding uncertainty $ b_{\mathrm{eff},PP}^{\mathrm{exp}} = 300 \ \mathrm{fT} $ are also considered \cite{schloss2018simultaneous}.
	The sensitivity can be improved by enlargement of the number of NV centers $ N $.
	However, there are many technical difficulties limit the enlargement of $ N $, such as synthesis of large-size single-crystal diamond and coherent control of NV centers in the large-scale diamond \cite{baryshev2023scalable,barry2020sensitivity,hopper2018spin}.

	The coupling parameter for $V_{PP}$ is
	\begin{equation}
		g_{P}^{e}g_{P}^{N}
		= \dfrac{b_{\mathrm{eff},PP}^{\mathrm{exp}}}{b_{\mathrm{eff},PP}^{\mathrm{sim}}|_{_{g_{P}^{e}g_{P}^{N}=1}}},\label{gg}
	\end{equation}
	where the magnetic polarization of the nuclear spins can be deduced from the strength of the magnetic dipole-dipole interaction $ (P_{+} - P_{-})_{\mathrm{exp}} = b_{\mathrm{dip}}^{\mathrm{exp}} / b_{\mathrm{dip}}^{\mathrm{sim}}|_{_{P_{+} - P_{-} = 100 \%}} $.

\begin{figure}[htp]
		\includegraphics[width=1\linewidth]{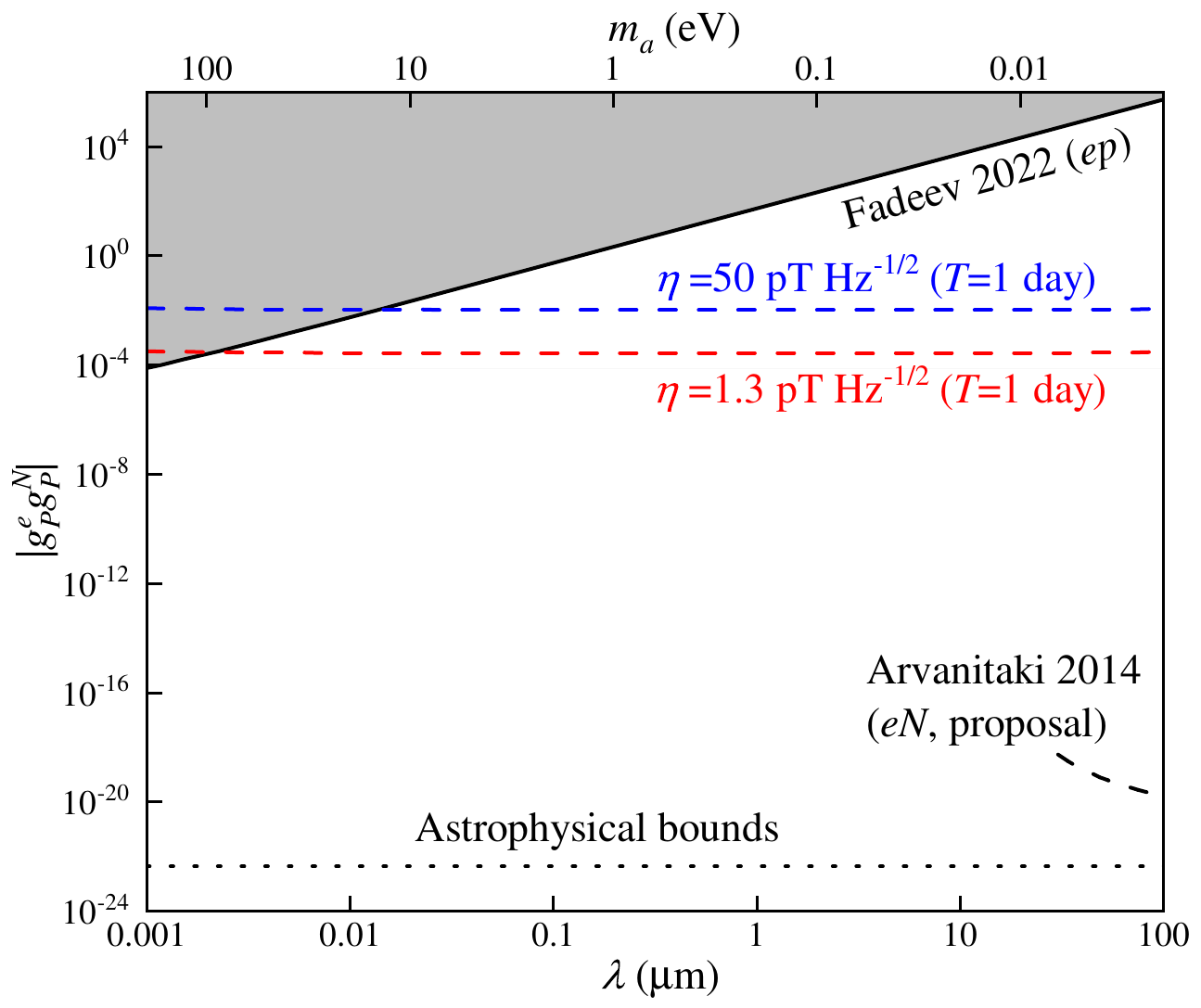}
		\caption{\label{fig:V3Limit} Upper limits on $ |g_{P}^{e} g_{P}^{N}| $ as a function of the force range $ \lambda $ and mass $ m_{a} $ of axions.
			The red and blue lines represent the expected constraints in this proposal with the total measurement time $ T = 1 \ \mathrm{day} $.
			The solid line represents the existing experimental constraint established by \cite{fadeev2022pseudovector}.
			Astrophysical bounds \cite{ohare2020cornering} and the expected constraint in a previous proposal \cite{arvanitaki2014resonantly} are also shown.
		}
	\end{figure}

\begin{figure}[H]
		\includegraphics[width=1\linewidth]{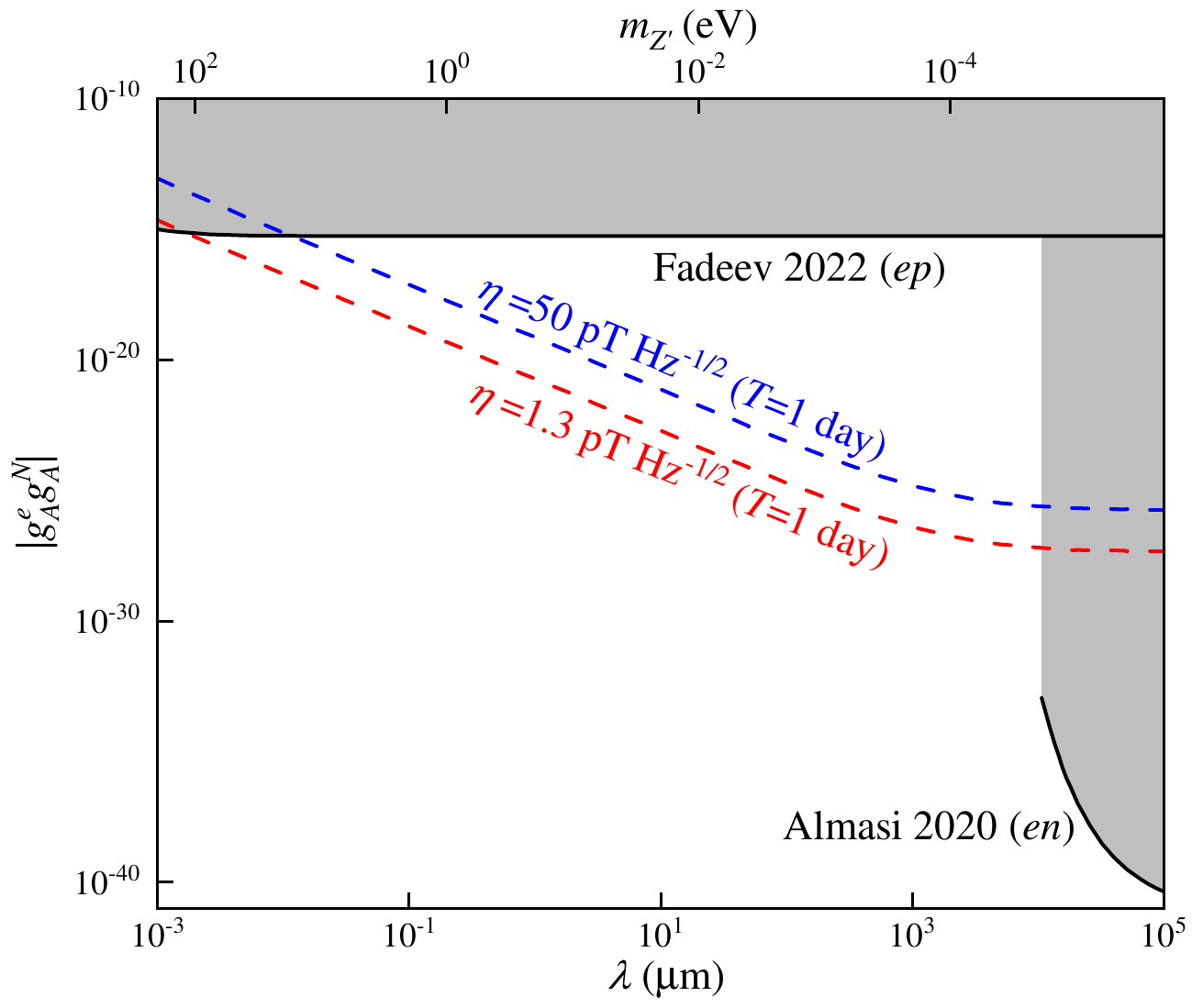}
		\caption{\label{fig:V2Limit} Upper limits on $ |g_{A}^{e} g_{A}^{N}| $ as a function of the force range $ \lambda $ and mass $ m_{Z'} $ of $ Z' $ bosons.
			The red and blue lines represent the expected constraints in this proposal with the total measurement time $ T = 1 \ \mathrm{day} $.
			The solid lines represent the existing experimental constraints established by \cite{almasi2020new,fadeev2022pseudovector}.
		}
	\end{figure}

	By adapting the parameters shown in Tab.~\ref{tab:parameters},
	the expected constraints on $ |g_{P}^{e}g_{P}^{N}| $ within the force range $\lambda$ from 1 nm to 100 $\upmu$m can be calculated,
	as shown in Fig.~\ref{fig:V3Limit}.
	The expected constraints are dominated by the last term in Eqs.~\ref{eqs:BeffPP}.
	The integration of factor $ e^{-r/\lambda} / (\lambda^{2}r) $ in this term remains constant, and results in the constant expected constraints.
	Similarly, the expected constraints on $ |g_{A}^{e}g_{A}^{N}| $ can also be obtained by replacing $ b_{\mathrm{eff},PP}^{\mathrm{sim}} $ by $ b_{\mathrm{eff},AA}^{\mathrm{sim}} $.
	The results are shown in Fig.~\ref{fig:V2Limit}.

	\begin{table}[h]
	\caption{\label{tab:parameters}
		Diamond geometry and other properties used in calculations.
	}
	\begin{ruledtabular}
		\begin{tabular}{lc}
			\textrm{Parameter}&\textrm{Value}\\
			\colrule
			Diamond thickness & 2 mm \\
			Diamond length (x direction) & 10 mm \\
			Diamond length (y direction) & 10 mm \\
			NV concentration & 0.1 ppm \\
			Nuclear magnetic polarization $ P_{+} - P_{-} $ & 70$\%$ \\
			Nucleon spin fraction of $ ^{14}\mathrm{N} $ $\zeta$ & 0.5 \\
			Sensitivity $ \eta $ & 1.3 pT/$ \sqrt{\mathrm{Hz}} $ \\

		\end{tabular}
	\end{ruledtabular}
	\end{table}

	\section{Conclusion \label{sec:Conclusion}}

	In conclusion, we propose an experimental search for the exotic interactions $ V_{PP} $ and $ V_{AA} $ between electrons and nucleons.
	NV centers along three different crystallographic axes of a single-crystal diamond can be utilized to construct the vector magnetometer.
	The $ ^{14}\mathrm{N} $ nuclear spins of NV centers along the fourth axis can be polarized through the DNP method as the spin sources.
	With a carefully designed external magnetic field, NV centers along different axes can be manipulated and read out separately.
	The effective magnetic fields induced by the exotic interactions can be distinguished from the dipole magnetic field by the different directions.
	In such a case, the nanocale distance from each spin sensor to the nearest spin source allows for exploring the force range at the nanometer scale.
	For the axion-mediated interaction $ V_{PP} $, new upper bounds of the coupling $ |g_{P}^{e} g_{P}^{N}| $ are expected within the force range from 10 nm to 100 $ \upmu $m.
	For the $ Z' $-mediated interaction $ V_{AA} $, new upper bounds of the coupling $ |g_{A}^{e} g_{A}^{N}| $ are expected within the force range from 10 nm to 1 cm.
	The new upper bounds for $ V_{PP} $ and $ V_{AA} $ are both expected to be more than 5 orders of magnitude more stringent than existing constraints at the force range of 1 $ \upmu $m with the total measurement time of one day.
	\\

	\begin{acknowledgments}
		This work was supported by NSFC (12150010, 12261160569, 12205290, T2388102) and the Innovation Program for Quantum Science and Technology (2021ZD0302200).
		X. R. thanks the Youth Innovation Promotion Association of Chinese Academy of Sciences for the support.
		A. L. G. and R. Q. K. contributed equally to this work.

	\end{acknowledgments}


\begin{thebibliography}{63}%
	\makeatletter
	\providecommand \@ifxundefined [1]{%
		\@ifx{#1\undefined}
	}%
	\providecommand \@ifnum [1]{%
		\ifnum #1\expandafter \@firstoftwo
		\else \expandafter \@secondoftwo
		\fi
	}%
	\providecommand \@ifx [1]{%
		\ifx #1\expandafter \@firstoftwo
		\else \expandafter \@secondoftwo
		\fi
	}%
	\providecommand \natexlab [1]{#1}%
	\providecommand \enquote  [1]{``#1''}%
	\providecommand \bibnamefont  [1]{#1}%
	\providecommand \bibfnamefont [1]{#1}%
	\providecommand \citenamefont [1]{#1}%
	\providecommand \href@noop [0]{\@secondoftwo}%
	\providecommand \href [0]{\begingroup \@sanitize@url \@href}%
	\providecommand \@href[1]{\@@startlink{#1}\@@href}%
	\providecommand \@@href[1]{\endgroup#1\@@endlink}%
	\providecommand \@sanitize@url [0]{\catcode `\\12\catcode `\$12\catcode
		`\&12\catcode `\#12\catcode `\^12\catcode `\_12\catcode `\%12\relax}%
	\providecommand \@@startlink[1]{}%
	\providecommand \@@endlink[0]{}%
	\providecommand \url  [0]{\begingroup\@sanitize@url \@url }%
	\providecommand \@url [1]{\endgroup\@href {#1}{\urlprefix }}%
	\providecommand \urlprefix  [0]{URL }%
	\providecommand \Eprint [0]{\href }%
	\providecommand \doibase [0]{https://doi.org/}%
	\providecommand \selectlanguage [0]{\@gobble}%
	\providecommand \bibinfo  [0]{\@secondoftwo}%
	\providecommand \bibfield  [0]{\@secondoftwo}%
	\providecommand \translation [1]{[#1]}%
	\providecommand \BibitemOpen [0]{}%
	\providecommand \bibitemStop [0]{}%
	\providecommand \bibitemNoStop [0]{.\EOS\space}%
	\providecommand \EOS [0]{\spacefactor3000\relax}%
	\providecommand \BibitemShut  [1]{\csname bibitem#1\endcsname}%
	\let\auto@bib@innerbib\@empty
	\bibitem [{\citenamefont {Aghanim}\ \emph {et~al.}(2020)\citenamefont {Aghanim}
		\emph {et~al.}}]{aghanim2020planck}%
	\BibitemOpen
	\bibfield  {author} {\bibinfo {author} {\bibfnamefont {N.}~\bibnamefont
			{Aghanim}} \emph {et~al.} (\bibinfo {collaboration} {{Planck
				Collaboration}}),\ }\bibfield  {title} {\bibinfo {title} {Planck 2018 results
			- {{VI}}. {{Cosmological}} parameters},\ }\href
	{https://doi.org/10.1051/0004-6361/201833910} {\bibfield  {journal} {\bibinfo
			{journal} {Astron. Astrophys.}\ }\textbf {\bibinfo {volume} {641}},\
		\bibinfo {pages} {A6} (\bibinfo {year} {2020})}\BibitemShut {NoStop}%
	\bibitem [{\citenamefont {Workman}\ \emph {et~al.}(2022)\citenamefont {Workman}
		\emph {et~al.}}]{particledatagroup2022review}%
	\BibitemOpen
	\bibfield  {author} {\bibinfo {author} {\bibfnamefont {R.~L.}\ \bibnamefont
			{Workman}} \emph {et~al.} (\bibinfo {collaboration} {{Particle Data
				Group}}),\ }\bibfield  {title} {\bibinfo {title} {Review of {{Particle
					Physics}}},\ }\href {https://doi.org/10.1093/ptep/ptac097} {\bibfield
		{journal} {\bibinfo  {journal} {Prog. Theor. Exp. Phys.}\ }\textbf {\bibinfo
			{volume} {2022}},\ \bibinfo {pages} {083C01} (\bibinfo {year}
		{2022})}\BibitemShut {NoStop}%
	\bibitem [{\citenamefont {Duffy}\ and\ \citenamefont {van
			Bibber}(2009)}]{duffy2009axions}%
	\BibitemOpen
	\bibfield  {author} {\bibinfo {author} {\bibfnamefont {L.~D.}\ \bibnamefont
			{Duffy}}\ and\ \bibinfo {author} {\bibfnamefont {K.}~\bibnamefont {van
				Bibber}},\ }\bibfield  {title} {\bibinfo {title} {Axions as dark matter
			particles},\ }\href {https://doi.org/10.1088/1367-2630/11/10/105008}
	{\bibfield  {journal} {\bibinfo  {journal} {New J. Phys.}\ }\textbf {\bibinfo
			{volume} {11}},\ \bibinfo {pages} {105008} (\bibinfo {year}
		{2009})}\BibitemShut {NoStop}%
	\bibitem [{\citenamefont {Borsanyi}\ \emph {et~al.}(2016)\citenamefont
		{Borsanyi}, \citenamefont {Fodor}, \citenamefont {Guenther}, \citenamefont
		{Kampert}, \citenamefont {Katz}, \citenamefont {Kawanai}, \citenamefont
		{Kovacs}, \citenamefont {Mages}, \citenamefont {Pasztor}, \citenamefont
		{Pittler}, \citenamefont {Redondo}, \citenamefont {Ringwald},\ and\
		\citenamefont {Szabo}}]{borsanyi2016calculation}%
	\BibitemOpen
	\bibfield  {author} {\bibinfo {author} {\bibfnamefont {S.}~\bibnamefont
			{Borsanyi}}, \bibinfo {author} {\bibfnamefont {Z.}~\bibnamefont {Fodor}},
		\bibinfo {author} {\bibfnamefont {J.}~\bibnamefont {Guenther}}, \bibinfo
		{author} {\bibfnamefont {K.-H.}\ \bibnamefont {Kampert}}, \bibinfo {author}
		{\bibfnamefont {S.~D.}\ \bibnamefont {Katz}}, \bibinfo {author}
		{\bibfnamefont {T.}~\bibnamefont {Kawanai}}, \bibinfo {author} {\bibfnamefont
			{T.~G.}\ \bibnamefont {Kovacs}}, \bibinfo {author} {\bibfnamefont {S.~W.}\
			\bibnamefont {Mages}}, \bibinfo {author} {\bibfnamefont {A.}~\bibnamefont
			{Pasztor}}, \bibinfo {author} {\bibfnamefont {F.}~\bibnamefont {Pittler}},
		\bibinfo {author} {\bibfnamefont {J.}~\bibnamefont {Redondo}}, \bibinfo
		{author} {\bibfnamefont {A.}~\bibnamefont {Ringwald}},\ and\ \bibinfo
		{author} {\bibfnamefont {K.~K.}\ \bibnamefont {Szabo}},\ }\bibfield  {title}
	{\bibinfo {title} {Calculation of the axion mass based on high-temperature
			lattice quantum chromodynamics},\ }\href
	{https://doi.org/10.1038/nature20115} {\bibfield  {journal} {\bibinfo
			{journal} {Nature}\ }\textbf {\bibinfo {volume} {539}},\ \bibinfo {pages}
		{69} (\bibinfo {year} {2016})}\BibitemShut {NoStop}%
	\bibitem [{\citenamefont {Dine}\ \emph {et~al.}(2017)\citenamefont {Dine},
		\citenamefont {Draper}, \citenamefont {{Stephenson-Haskins}},\ and\
		\citenamefont {Xu}}]{dine2017axions}%
	\BibitemOpen
	\bibfield  {author} {\bibinfo {author} {\bibfnamefont {M.}~\bibnamefont
			{Dine}}, \bibinfo {author} {\bibfnamefont {P.}~\bibnamefont {Draper}},
		\bibinfo {author} {\bibfnamefont {L.}~\bibnamefont {{Stephenson-Haskins}}},\
		and\ \bibinfo {author} {\bibfnamefont {D.}~\bibnamefont {Xu}},\ }\bibfield
	{title} {\bibinfo {title} {Axions, instantons, and the lattice},\ }\href
	{https://doi.org/10.1103/PhysRevD.96.095001} {\bibfield  {journal} {\bibinfo
			{journal} {Phys. Rev. D}\ }\textbf {\bibinfo {volume} {96}},\ \bibinfo
		{pages} {095001} (\bibinfo {year} {2017})}\BibitemShut {NoStop}%
	\bibitem [{\citenamefont {Ballesteros}\ \emph {et~al.}(2017)\citenamefont
		{Ballesteros}, \citenamefont {Redondo}, \citenamefont {Ringwald},\ and\
		\citenamefont {Tamarit}}]{ballesteros2017unifying}%
	\BibitemOpen
	\bibfield  {author} {\bibinfo {author} {\bibfnamefont {G.}~\bibnamefont
			{Ballesteros}}, \bibinfo {author} {\bibfnamefont {J.}~\bibnamefont
			{Redondo}}, \bibinfo {author} {\bibfnamefont {A.}~\bibnamefont {Ringwald}},\
		and\ \bibinfo {author} {\bibfnamefont {C.}~\bibnamefont {Tamarit}},\
	}\bibfield  {title} {\bibinfo {title} {Unifying {{Inflation}} with the
			{{Axion}}, {{Dark Matter}}, {{Baryogenesis}}, and the {{Seesaw Mechanism}}},\
	}\href {https://doi.org/10.1103/PhysRevLett.118.071802} {\bibfield  {journal}
		{\bibinfo  {journal} {Phys. Rev. Lett.}\ }\textbf {\bibinfo {volume} {118}},\
		\bibinfo {pages} {071802} (\bibinfo {year} {2017})}\BibitemShut {NoStop}%
	\bibitem [{\citenamefont {{Chadha-Day}}\ \emph {et~al.}(2022)\citenamefont
		{{Chadha-Day}}, \citenamefont {Ellis},\ and\ \citenamefont
		{Marsh}}]{chadha-day2022axion}%
	\BibitemOpen
	\bibfield  {author} {\bibinfo {author} {\bibfnamefont {F.}~\bibnamefont
			{{Chadha-Day}}}, \bibinfo {author} {\bibfnamefont {J.}~\bibnamefont
			{Ellis}},\ and\ \bibinfo {author} {\bibfnamefont {D.~J.~E.}\ \bibnamefont
			{Marsh}},\ }\bibfield  {title} {\bibinfo {title} {Axion dark matter: {{What}}
			is it and why now?},\ }\href {https://doi.org/10.1126/sciadv.abj3618}
	{\bibfield  {journal} {\bibinfo  {journal} {Sci. Adv.}\ }\textbf {\bibinfo
			{volume} {8}},\ \bibinfo {pages} {eabj3618} (\bibinfo {year}
		{2022})}\BibitemShut {NoStop}%
	\bibitem [{\citenamefont {Peccei}\ and\ \citenamefont
		{Quinn}(1977{\natexlab{a}})}]{peccei1977constraints}%
	\BibitemOpen
	\bibfield  {author} {\bibinfo {author} {\bibfnamefont {R.~D.}\ \bibnamefont
			{Peccei}}\ and\ \bibinfo {author} {\bibfnamefont {H.~R.}\ \bibnamefont
			{Quinn}},\ }\bibfield  {title} {\bibinfo {title} {Constraints imposed by
			$\mathrm{CP}$ conservation in the presence of pseudoparticles},\ }\href
	{https://doi.org/10.1103/PhysRevD.16.1791} {\bibfield  {journal} {\bibinfo
			{journal} {Phys. Rev. D}\ }\textbf {\bibinfo {volume} {16}},\ \bibinfo
		{pages} {1791} (\bibinfo {year} {1977}{\natexlab{a}})}\BibitemShut {NoStop}%
	\bibitem [{\citenamefont {Peccei}\ and\ \citenamefont
		{Quinn}(1977{\natexlab{b}})}]{peccei1977mathrm}%
	\BibitemOpen
	\bibfield  {author} {\bibinfo {author} {\bibfnamefont {R.~D.}\ \bibnamefont
			{Peccei}}\ and\ \bibinfo {author} {\bibfnamefont {H.~R.}\ \bibnamefont
			{Quinn}},\ }\bibfield  {title} {\bibinfo {title} {$\mathrm{CP}$
			{{Conservation}} in the {{Presence}} of {{Pseudoparticles}}},\ }\href
	{https://doi.org/10.1103/PhysRevLett.38.1440} {\bibfield  {journal} {\bibinfo
			{journal} {Phys. Rev. Lett.}\ }\textbf {\bibinfo {volume} {38}},\ \bibinfo
		{pages} {1440} (\bibinfo {year} {1977}{\natexlab{b}})}\BibitemShut {NoStop}%
	\bibitem [{\citenamefont {Weinberg}(1978)}]{weinberg1978new}%
	\BibitemOpen
	\bibfield  {author} {\bibinfo {author} {\bibfnamefont {S.}~\bibnamefont
			{Weinberg}},\ }\bibfield  {title} {\bibinfo {title} {A {{New Light
					Boson}}?},\ }\href {https://doi.org/10.1103/PhysRevLett.40.223} {\bibfield
		{journal} {\bibinfo  {journal} {Phys. Rev. Lett.}\ }\textbf {\bibinfo
			{volume} {40}},\ \bibinfo {pages} {223} (\bibinfo {year} {1978})}\BibitemShut
	{NoStop}%
	\bibitem [{\citenamefont {Wilczek}(1978)}]{wilczek1978problem}%
	\BibitemOpen
	\bibfield  {author} {\bibinfo {author} {\bibfnamefont {F.}~\bibnamefont
			{Wilczek}},\ }\bibfield  {title} {\bibinfo {title} {Problem of {{Strong}}
			${{P}}$ and ${{T}}$ {{Invariance}} in the {{Presence}} of {{Instantons}}},\
	}\href {https://doi.org/10.1103/PhysRevLett.40.279} {\bibfield  {journal}
		{\bibinfo  {journal} {Phys. Rev. Lett.}\ }\textbf {\bibinfo {volume} {40}},\
		\bibinfo {pages} {279} (\bibinfo {year} {1978})}\BibitemShut {NoStop}%
	\bibitem [{\citenamefont {Moody}\ and\ \citenamefont
		{Wilczek}(1984)}]{moody1984new}%
	\BibitemOpen
	\bibfield  {author} {\bibinfo {author} {\bibfnamefont {J.~E.}\ \bibnamefont
			{Moody}}\ and\ \bibinfo {author} {\bibfnamefont {F.}~\bibnamefont
			{Wilczek}},\ }\bibfield  {title} {\bibinfo {title} {New macroscopic
			forces?},\ }\href {https://doi.org/10.1103/PhysRevD.30.130} {\bibfield
		{journal} {\bibinfo  {journal} {Phys. Rev. D}\ }\textbf {\bibinfo {volume}
			{30}},\ \bibinfo {pages} {130} (\bibinfo {year} {1984})}\BibitemShut
	{NoStop}%
	\bibitem [{\citenamefont {Dobrescu}\ and\ \citenamefont
		{Mocioiu}(2006)}]{dobrescu2006spindependent}%
	\BibitemOpen
	\bibfield  {author} {\bibinfo {author} {\bibfnamefont {B.~A.}\ \bibnamefont
			{Dobrescu}}\ and\ \bibinfo {author} {\bibfnamefont {I.}~\bibnamefont
			{Mocioiu}},\ }\bibfield  {title} {\bibinfo {title} {Spin-dependent
			macroscopic forces from new particle exchange},\ }\href
	{https://doi.org/10.1088/1126-6708/2006/11/005} {\bibfield  {journal}
		{\bibinfo  {journal} {J. High Energy Phys.}\ }\textbf {\bibinfo {volume}
			{2006}}\bibinfo  {number} { (11)},\ \bibinfo {pages} {005}}\BibitemShut
	{NoStop}%
	\bibitem [{\citenamefont {Fadeev}\ \emph {et~al.}(2019)\citenamefont {Fadeev},
		\citenamefont {Stadnik}, \citenamefont {Ficek}, \citenamefont {Kozlov},
		\citenamefont {Flambaum},\ and\ \citenamefont
		{Budker}}]{fadeev2019revisiting}%
	\BibitemOpen
	\bibfield  {number} {  }\bibfield  {author} {\bibinfo {author} {\bibfnamefont
			{P.}~\bibnamefont {Fadeev}}, \bibinfo {author} {\bibfnamefont {Y.~V.}\
			\bibnamefont {Stadnik}}, \bibinfo {author} {\bibfnamefont {F.}~\bibnamefont
			{Ficek}}, \bibinfo {author} {\bibfnamefont {M.~G.}\ \bibnamefont {Kozlov}},
		\bibinfo {author} {\bibfnamefont {V.~V.}\ \bibnamefont {Flambaum}},\ and\
		\bibinfo {author} {\bibfnamefont {D.}~\bibnamefont {Budker}},\ }\bibfield
	{title} {\bibinfo {title} {Revisiting spin-dependent forces mediated by new
			bosons: {{Potentials}} in the coordinate-space representation for
			macroscopic- and atomic-scale experiments},\ }\href
	{https://doi.org/10.1103/PhysRevA.99.022113} {\bibfield  {journal} {\bibinfo
			{journal} {Phys. Rev. A}\ }\textbf {\bibinfo {volume} {99}},\ \bibinfo
		{pages} {022113} (\bibinfo {year} {2019})}\BibitemShut {NoStop}%
	\bibitem [{\citenamefont {Safronova}\ \emph {et~al.}(2018)\citenamefont
		{Safronova}, \citenamefont {Budker}, \citenamefont {DeMille}, \citenamefont
		{Kimball}, \citenamefont {Derevianko},\ and\ \citenamefont
		{Clark}}]{safronova2018search}%
	\BibitemOpen
	\bibfield  {author} {\bibinfo {author} {\bibfnamefont {M.~S.}\ \bibnamefont
			{Safronova}}, \bibinfo {author} {\bibfnamefont {D.}~\bibnamefont {Budker}},
		\bibinfo {author} {\bibfnamefont {D.}~\bibnamefont {DeMille}}, \bibinfo
		{author} {\bibfnamefont {D.~F.~J.}\ \bibnamefont {Kimball}}, \bibinfo
		{author} {\bibfnamefont {A.}~\bibnamefont {Derevianko}},\ and\ \bibinfo
		{author} {\bibfnamefont {C.~W.}\ \bibnamefont {Clark}},\ }\bibfield  {title}
	{\bibinfo {title} {Search for new physics with atoms and molecules},\ }\href
	{https://doi.org/10.1103/RevModPhys.90.025008} {\bibfield  {journal}
		{\bibinfo  {journal} {Rev. Mod. Phys.}\ }\textbf {\bibinfo {volume} {90}},\
		\bibinfo {pages} {025008} (\bibinfo {year} {2018})}\BibitemShut {NoStop}%
	\bibitem [{\citenamefont {Terrano}\ \emph {et~al.}(2015)\citenamefont
		{Terrano}, \citenamefont {Adelberger}, \citenamefont {Lee},\ and\
		\citenamefont {Heckel}}]{terrano2015shortrange}%
	\BibitemOpen
	\bibfield  {author} {\bibinfo {author} {\bibfnamefont {W.~A.}\ \bibnamefont
			{Terrano}}, \bibinfo {author} {\bibfnamefont {E.~G.}\ \bibnamefont
			{Adelberger}}, \bibinfo {author} {\bibfnamefont {J.~G.}\ \bibnamefont
			{Lee}},\ and\ \bibinfo {author} {\bibfnamefont {B.~R.}\ \bibnamefont
			{Heckel}},\ }\bibfield  {title} {\bibinfo {title} {Short-{{Range}},
			{{Spin-Dependent Interactions}} of {{Electrons}}: {{A Probe}} for {{Exotic
					Pseudo-Goldstone Bosons}}},\ }\href
	{https://doi.org/10.1103/PhysRevLett.115.201801} {\bibfield  {journal}
		{\bibinfo  {journal} {Phys. Rev. Lett.}\ }\textbf {\bibinfo {volume} {115}},\
		\bibinfo {pages} {201801} (\bibinfo {year} {2015})}\BibitemShut {NoStop}%
	\bibitem [{\citenamefont {Hoedl}\ \emph {et~al.}(2011)\citenamefont {Hoedl},
		\citenamefont {Fleischer}, \citenamefont {Adelberger},\ and\ \citenamefont
		{Heckel}}]{hoedl2011improved}%
	\BibitemOpen
	\bibfield  {author} {\bibinfo {author} {\bibfnamefont {S.~A.}\ \bibnamefont
			{Hoedl}}, \bibinfo {author} {\bibfnamefont {S.~M.}\ \bibnamefont
			{Fleischer}}, \bibinfo {author} {\bibfnamefont {E.~G.}\ \bibnamefont
			{Adelberger}},\ and\ \bibinfo {author} {\bibfnamefont {B.~R.}\ \bibnamefont
			{Heckel}},\ }\bibfield  {title} {\bibinfo {title} {Improved {{Constraints}}
			on an {{Axion-Mediated Force}}},\ }\href
	{https://doi.org/10.1103/PhysRevLett.106.041801} {\bibfield  {journal}
		{\bibinfo  {journal} {Phys. Rev. Lett.}\ }\textbf {\bibinfo {volume} {106}},\
		\bibinfo {pages} {041801} (\bibinfo {year} {2011})}\BibitemShut {NoStop}%
	\bibitem [{\citenamefont {Heckel}\ \emph {et~al.}(2008)\citenamefont {Heckel},
		\citenamefont {Adelberger}, \citenamefont {Cramer}, \citenamefont {Cook},
		\citenamefont {Schlamminger},\ and\ \citenamefont
		{Schmidt}}]{heckel2008preferredframe}%
	\BibitemOpen
	\bibfield  {author} {\bibinfo {author} {\bibfnamefont {B.~R.}\ \bibnamefont
			{Heckel}}, \bibinfo {author} {\bibfnamefont {E.~G.}\ \bibnamefont
			{Adelberger}}, \bibinfo {author} {\bibfnamefont {C.~E.}\ \bibnamefont
			{Cramer}}, \bibinfo {author} {\bibfnamefont {T.~S.}\ \bibnamefont {Cook}},
		\bibinfo {author} {\bibfnamefont {S.}~\bibnamefont {Schlamminger}},\ and\
		\bibinfo {author} {\bibfnamefont {U.}~\bibnamefont {Schmidt}},\ }\bibfield
	{title} {\bibinfo {title} {Preferred-frame and $\mathrm{CP}$-violation tests
			with polarized electrons},\ }\href
	{https://doi.org/10.1103/PhysRevD.78.092006} {\bibfield  {journal} {\bibinfo
			{journal} {Phys. Rev. D}\ }\textbf {\bibinfo {volume} {78}},\ \bibinfo
		{pages} {092006} (\bibinfo {year} {2008})}\BibitemShut {NoStop}%
	\bibitem [{\citenamefont {Ritter}\ \emph {et~al.}(1990)\citenamefont {Ritter},
		\citenamefont {Goldblum}, \citenamefont {Ni}, \citenamefont {Gillies},\ and\
		\citenamefont {Speake}}]{ritter1990experimental}%
	\BibitemOpen
	\bibfield  {author} {\bibinfo {author} {\bibfnamefont {R.~C.}\ \bibnamefont
			{Ritter}}, \bibinfo {author} {\bibfnamefont {C.~E.}\ \bibnamefont
			{Goldblum}}, \bibinfo {author} {\bibfnamefont {W.-T.}\ \bibnamefont {Ni}},
		\bibinfo {author} {\bibfnamefont {G.~T.}\ \bibnamefont {Gillies}},\ and\
		\bibinfo {author} {\bibfnamefont {C.~C.}\ \bibnamefont {Speake}},\ }\bibfield
	{title} {\bibinfo {title} {Experimental test of equivalence principle with
			polarized masses},\ }\href {https://doi.org/10.1103/PhysRevD.42.977}
	{\bibfield  {journal} {\bibinfo  {journal} {Phys. Rev. D}\ }\textbf {\bibinfo
			{volume} {42}},\ \bibinfo {pages} {977} (\bibinfo {year} {1990})}\BibitemShut
	{NoStop}%
	\bibitem [{\citenamefont {Kotler}\ \emph {et~al.}(2015)\citenamefont {Kotler},
		\citenamefont {Ozeri},\ and\ \citenamefont
		{Kimball}}]{kotler2015constraints}%
	\BibitemOpen
	\bibfield  {author} {\bibinfo {author} {\bibfnamefont {S.}~\bibnamefont
			{Kotler}}, \bibinfo {author} {\bibfnamefont {R.}~\bibnamefont {Ozeri}},\ and\
		\bibinfo {author} {\bibfnamefont {D.~F.~J.}\ \bibnamefont {Kimball}},\
	}\bibfield  {title} {\bibinfo {title} {Constraints on {{Exotic Dipole-Dipole
					Couplings}} between {{Electrons}} at the {{Micrometer Scale}}},\ }\href
	{https://doi.org/10.1103/PhysRevLett.115.081801} {\bibfield  {journal}
		{\bibinfo  {journal} {Phys. Rev. Lett.}\ }\textbf {\bibinfo {volume} {115}},\
		\bibinfo {pages} {081801} (\bibinfo {year} {2015})}\BibitemShut {NoStop}%
	\bibitem [{\citenamefont {Ding}\ \emph {et~al.}(2020)\citenamefont {Ding},
		\citenamefont {Wang}, \citenamefont {Zhou}, \citenamefont {Liu},
		\citenamefont {Sun}, \citenamefont {Adeyeye}, \citenamefont {Fu},
		\citenamefont {Ren}, \citenamefont {Li}, \citenamefont {Luo}, \citenamefont
		{Lan}, \citenamefont {Yang},\ and\ \citenamefont
		{Luo}}]{ding2020constraints}%
	\BibitemOpen
	\bibfield  {author} {\bibinfo {author} {\bibfnamefont {J.}~\bibnamefont
			{Ding}}, \bibinfo {author} {\bibfnamefont {J.}~\bibnamefont {Wang}}, \bibinfo
		{author} {\bibfnamefont {X.}~\bibnamefont {Zhou}}, \bibinfo {author}
		{\bibfnamefont {Y.}~\bibnamefont {Liu}}, \bibinfo {author} {\bibfnamefont
			{K.}~\bibnamefont {Sun}}, \bibinfo {author} {\bibfnamefont {A.~O.}\
			\bibnamefont {Adeyeye}}, \bibinfo {author} {\bibfnamefont {H.}~\bibnamefont
			{Fu}}, \bibinfo {author} {\bibfnamefont {X.}~\bibnamefont {Ren}}, \bibinfo
		{author} {\bibfnamefont {S.}~\bibnamefont {Li}}, \bibinfo {author}
		{\bibfnamefont {P.}~\bibnamefont {Luo}}, \bibinfo {author} {\bibfnamefont
			{Z.}~\bibnamefont {Lan}}, \bibinfo {author} {\bibfnamefont {S.}~\bibnamefont
			{Yang}},\ and\ \bibinfo {author} {\bibfnamefont {J.}~\bibnamefont {Luo}},\
	}\bibfield  {title} {\bibinfo {title} {Constraints on the {{Velocity}} and
			{{Spin Dependent Exotic Interaction}} at the {{Micrometer Range}}},\ }\href
	{https://doi.org/10.1103/PhysRevLett.124.161801} {\bibfield  {journal}
		{\bibinfo  {journal} {Phys. Rev. Lett.}\ }\textbf {\bibinfo {volume} {124}},\
		\bibinfo {pages} {161801} (\bibinfo {year} {2020})}\BibitemShut {NoStop}%
	\bibitem [{\citenamefont {Ficek}\ \emph {et~al.}(2018)\citenamefont {Ficek},
		\citenamefont {Fadeev}, \citenamefont {Flambaum}, \citenamefont
		{Jackson~Kimball}, \citenamefont {Kozlov}, \citenamefont {Stadnik},\ and\
		\citenamefont {Budker}}]{ficek2018constraints}%
	\BibitemOpen
	\bibfield  {author} {\bibinfo {author} {\bibfnamefont {F.}~\bibnamefont
			{Ficek}}, \bibinfo {author} {\bibfnamefont {P.}~\bibnamefont {Fadeev}},
		\bibinfo {author} {\bibfnamefont {V.~V.}\ \bibnamefont {Flambaum}}, \bibinfo
		{author} {\bibfnamefont {D.~F.}\ \bibnamefont {Jackson~Kimball}}, \bibinfo
		{author} {\bibfnamefont {M.~G.}\ \bibnamefont {Kozlov}}, \bibinfo {author}
		{\bibfnamefont {Y.~V.}\ \bibnamefont {Stadnik}},\ and\ \bibinfo {author}
		{\bibfnamefont {D.}~\bibnamefont {Budker}},\ }\bibfield  {title} {\bibinfo
		{title} {Constraints on {{Exotic Spin-Dependent Interactions Between Matter}}
			and {{Antimatter}} from {{Antiprotonic Helium Spectroscopy}}},\ }\href
	{https://doi.org/10.1103/PhysRevLett.120.183002} {\bibfield  {journal}
		{\bibinfo  {journal} {Phys. Rev. Lett.}\ }\textbf {\bibinfo {volume} {120}},\
		\bibinfo {pages} {183002} (\bibinfo {year} {2018})}\BibitemShut {NoStop}%
	\bibitem [{\citenamefont {Ficek}\ \emph {et~al.}(2017)\citenamefont {Ficek},
		\citenamefont {Kimball}, \citenamefont {Kozlov}, \citenamefont {Leefer},
		\citenamefont {Pustelny},\ and\ \citenamefont
		{Budker}}]{ficek2017constraints}%
	\BibitemOpen
	\bibfield  {author} {\bibinfo {author} {\bibfnamefont {F.}~\bibnamefont
			{Ficek}}, \bibinfo {author} {\bibfnamefont {D.~F.~J.}\ \bibnamefont
			{Kimball}}, \bibinfo {author} {\bibfnamefont {M.~G.}\ \bibnamefont {Kozlov}},
		\bibinfo {author} {\bibfnamefont {N.}~\bibnamefont {Leefer}}, \bibinfo
		{author} {\bibfnamefont {S.}~\bibnamefont {Pustelny}},\ and\ \bibinfo
		{author} {\bibfnamefont {D.}~\bibnamefont {Budker}},\ }\bibfield  {title}
	{\bibinfo {title} {Constraints on exotic spin-dependent interactions between
			electrons from helium fine-structure spectroscopy},\ }\href
	{https://doi.org/10.1103/PhysRevA.95.032505} {\bibfield  {journal} {\bibinfo
			{journal} {Phys. Rev. A}\ }\textbf {\bibinfo {volume} {95}},\ \bibinfo
		{pages} {032505} (\bibinfo {year} {2017})}\BibitemShut {NoStop}%
	\bibitem [{\citenamefont {Wu}\ \emph {et~al.}(2022)\citenamefont {Wu},
		\citenamefont {Chen}, \citenamefont {Sun}, \citenamefont {Peng},
		\citenamefont {Peng},\ and\ \citenamefont {Yan}}]{wu2022experimental}%
	\BibitemOpen
	\bibfield  {author} {\bibinfo {author} {\bibfnamefont {K.~Y.}\ \bibnamefont
			{Wu}}, \bibinfo {author} {\bibfnamefont {S.~Y.}\ \bibnamefont {Chen}},
		\bibinfo {author} {\bibfnamefont {G.~A.}\ \bibnamefont {Sun}}, \bibinfo
		{author} {\bibfnamefont {S.~M.}\ \bibnamefont {Peng}}, \bibinfo {author}
		{\bibfnamefont {M.}~\bibnamefont {Peng}},\ and\ \bibinfo {author}
		{\bibfnamefont {H.}~\bibnamefont {Yan}},\ }\bibfield  {title} {\bibinfo
		{title} {Experimental {{Limits}} on {{Exotic Spin}} and {{Velocity Dependent
					Interactions Using Rotationally Modulated Source Masses}} and an
			{{Atomic-Magnetometer Array}}},\ }\href
	{https://doi.org/10.1103/PhysRevLett.129.051802} {\bibfield  {journal}
		{\bibinfo  {journal} {Phys. Rev. Lett.}\ }\textbf {\bibinfo {volume} {129}},\
		\bibinfo {pages} {051802} (\bibinfo {year} {2022})}\BibitemShut {NoStop}%
	\bibitem [{\citenamefont {Almasi}\ \emph {et~al.}(2020)\citenamefont {Almasi},
		\citenamefont {Lee}, \citenamefont {Winarto}, \citenamefont {Smiciklas},\
		and\ \citenamefont {Romalis}}]{almasi2020new}%
	\BibitemOpen
	\bibfield  {author} {\bibinfo {author} {\bibfnamefont {A.}~\bibnamefont
			{Almasi}}, \bibinfo {author} {\bibfnamefont {J.}~\bibnamefont {Lee}},
		\bibinfo {author} {\bibfnamefont {H.}~\bibnamefont {Winarto}}, \bibinfo
		{author} {\bibfnamefont {M.}~\bibnamefont {Smiciklas}},\ and\ \bibinfo
		{author} {\bibfnamefont {M.~V.}\ \bibnamefont {Romalis}},\ }\bibfield
	{title} {\bibinfo {title} {New {{Limits}} on {{Anomalous Spin-Spin
					Interactions}}},\ }\href {https://doi.org/10.1103/PhysRevLett.125.201802}
	{\bibfield  {journal} {\bibinfo  {journal} {Phys. Rev. Lett.}\ }\textbf
		{\bibinfo {volume} {125}},\ \bibinfo {pages} {201802} (\bibinfo {year}
		{2020})}\BibitemShut {NoStop}%
	\bibitem [{\citenamefont {Kim}\ \emph {et~al.}(2019)\citenamefont {Kim},
		\citenamefont {Chu}, \citenamefont {Savukov},\ and\ \citenamefont
		{Newman}}]{kim2019experimental}%
	\BibitemOpen
	\bibfield  {author} {\bibinfo {author} {\bibfnamefont {Y.~J.}\ \bibnamefont
			{Kim}}, \bibinfo {author} {\bibfnamefont {P.-H.}\ \bibnamefont {Chu}},
		\bibinfo {author} {\bibfnamefont {I.}~\bibnamefont {Savukov}},\ and\ \bibinfo
		{author} {\bibfnamefont {S.}~\bibnamefont {Newman}},\ }\bibfield  {title}
	{\bibinfo {title} {Experimental limit on an exotic parity-odd spin- and
			velocity-dependent interaction using an optically polarized vapor},\ }\href
	{https://doi.org/10.1038/s41467-019-10169-1} {\bibfield  {journal} {\bibinfo
			{journal} {Nat. Commun.}\ }\textbf {\bibinfo {volume} {10}},\ \bibinfo
		{pages} {2245} (\bibinfo {year} {2019})}\BibitemShut {NoStop}%
	\bibitem [{\citenamefont {Kim}\ \emph {et~al.}(2018)\citenamefont {Kim},
		\citenamefont {Chu},\ and\ \citenamefont {Savukov}}]{kim2018experimental}%
	\BibitemOpen
	\bibfield  {author} {\bibinfo {author} {\bibfnamefont {Y.~J.}\ \bibnamefont
			{Kim}}, \bibinfo {author} {\bibfnamefont {P.-H.}\ \bibnamefont {Chu}},\ and\
		\bibinfo {author} {\bibfnamefont {I.}~\bibnamefont {Savukov}},\ }\bibfield
	{title} {\bibinfo {title} {Experimental {{Constraint}} on an {{Exotic Spin-}}
			and {{Velocity-Dependent Interaction}} in the {{Sub-meV Range}} of {{Axion
					Mass}} with a {{Spin-Exchange Relaxation-Free Magnetometer}}},\ }\href
	{https://doi.org/10.1103/PhysRevLett.121.091802} {\bibfield  {journal}
		{\bibinfo  {journal} {Phys. Rev. Lett.}\ }\textbf {\bibinfo {volume} {121}},\
		\bibinfo {pages} {091802} (\bibinfo {year} {2018})}\BibitemShut {NoStop}%
	\bibitem [{\citenamefont {Lee}\ \emph {et~al.}(2018)\citenamefont {Lee},
		\citenamefont {Almasi},\ and\ \citenamefont {Romalis}}]{lee2018improved}%
	\BibitemOpen
	\bibfield  {author} {\bibinfo {author} {\bibfnamefont {J.}~\bibnamefont
			{Lee}}, \bibinfo {author} {\bibfnamefont {A.}~\bibnamefont {Almasi}},\ and\
		\bibinfo {author} {\bibfnamefont {M.}~\bibnamefont {Romalis}},\ }\bibfield
	{title} {\bibinfo {title} {Improved {{Limits}} on {{Spin-Mass
					Interactions}}},\ }\href {https://doi.org/10.1103/PhysRevLett.120.161801}
	{\bibfield  {journal} {\bibinfo  {journal} {Phys. Rev. Lett.}\ }\textbf
		{\bibinfo {volume} {120}},\ \bibinfo {pages} {161801} (\bibinfo {year}
		{2018})}\BibitemShut {NoStop}%
	\bibitem [{\citenamefont {Ji}\ \emph {et~al.}(2018)\citenamefont {Ji},
		\citenamefont {Chen}, \citenamefont {Fu}, \citenamefont {Ding}, \citenamefont
		{Fang}, \citenamefont {Xiao}, \citenamefont {Wei},\ and\ \citenamefont
		{Yan}}]{ji2018new}%
	\BibitemOpen
	\bibfield  {author} {\bibinfo {author} {\bibfnamefont {W.}~\bibnamefont
			{Ji}}, \bibinfo {author} {\bibfnamefont {Y.}~\bibnamefont {Chen}}, \bibinfo
		{author} {\bibfnamefont {C.}~\bibnamefont {Fu}}, \bibinfo {author}
		{\bibfnamefont {M.}~\bibnamefont {Ding}}, \bibinfo {author} {\bibfnamefont
			{J.}~\bibnamefont {Fang}}, \bibinfo {author} {\bibfnamefont {Z.}~\bibnamefont
			{Xiao}}, \bibinfo {author} {\bibfnamefont {K.}~\bibnamefont {Wei}},\ and\
		\bibinfo {author} {\bibfnamefont {H.}~\bibnamefont {Yan}},\ }\bibfield
	{title} {\bibinfo {title} {New {{Experimental Limits}} on {{Exotic
					Spin-Spin-Velocity-Dependent Interactions}} by {{Using}}
			{${\mathrm{SmCo}}_{5}$} {{Spin Sources}}},\ }\href
	{https://doi.org/10.1103/PhysRevLett.121.261803} {\bibfield  {journal}
		{\bibinfo  {journal} {Phys. Rev. Lett.}\ }\textbf {\bibinfo {volume} {121}},\
		\bibinfo {pages} {261803} (\bibinfo {year} {2018})}\BibitemShut {NoStop}%
	\bibitem [{\citenamefont {Wang}\ \emph {et~al.}(2022)\citenamefont {Wang},
		\citenamefont {Su}, \citenamefont {Jiang}, \citenamefont {Huang},
		\citenamefont {Qin}, \citenamefont {Guo}, \citenamefont {Wang}, \citenamefont
		{Hu}, \citenamefont {Ji}, \citenamefont {Fadeev}, \citenamefont {Peng},\ and\
		\citenamefont {Budker}}]{wang2022limits}%
	\BibitemOpen
	\bibfield  {author} {\bibinfo {author} {\bibfnamefont {Y.}~\bibnamefont
			{Wang}}, \bibinfo {author} {\bibfnamefont {H.}~\bibnamefont {Su}}, \bibinfo
		{author} {\bibfnamefont {M.}~\bibnamefont {Jiang}}, \bibinfo {author}
		{\bibfnamefont {Y.}~\bibnamefont {Huang}}, \bibinfo {author} {\bibfnamefont
			{Y.}~\bibnamefont {Qin}}, \bibinfo {author} {\bibfnamefont {C.}~\bibnamefont
			{Guo}}, \bibinfo {author} {\bibfnamefont {Z.}~\bibnamefont {Wang}}, \bibinfo
		{author} {\bibfnamefont {D.}~\bibnamefont {Hu}}, \bibinfo {author}
		{\bibfnamefont {W.}~\bibnamefont {Ji}}, \bibinfo {author} {\bibfnamefont
			{P.}~\bibnamefont {Fadeev}}, \bibinfo {author} {\bibfnamefont
			{X.}~\bibnamefont {Peng}},\ and\ \bibinfo {author} {\bibfnamefont
			{D.}~\bibnamefont {Budker}},\ }\bibfield  {title} {\bibinfo {title} {Limits
			on {{Axions}} and {{Axionlike Particles}} within the {{Axion Window Using}} a
			{{Spin-Based Amplifier}}},\ }\href
	{https://doi.org/10.1103/PhysRevLett.129.051801} {\bibfield  {journal}
		{\bibinfo  {journal} {Phys. Rev. Lett.}\ }\textbf {\bibinfo {volume} {129}},\
		\bibinfo {pages} {051801} (\bibinfo {year} {2022})}\BibitemShut {NoStop}%
	\bibitem [{\citenamefont {Su}\ \emph {et~al.}(2021)\citenamefont {Su},
		\citenamefont {Wang}, \citenamefont {Jiang}, \citenamefont {Ji},
		\citenamefont {Fadeev}, \citenamefont {Hu}, \citenamefont {Peng},\ and\
		\citenamefont {Budker}}]{su2021search}%
	\BibitemOpen
	\bibfield  {author} {\bibinfo {author} {\bibfnamefont {H.}~\bibnamefont
			{Su}}, \bibinfo {author} {\bibfnamefont {Y.}~\bibnamefont {Wang}}, \bibinfo
		{author} {\bibfnamefont {M.}~\bibnamefont {Jiang}}, \bibinfo {author}
		{\bibfnamefont {W.}~\bibnamefont {Ji}}, \bibinfo {author} {\bibfnamefont
			{P.}~\bibnamefont {Fadeev}}, \bibinfo {author} {\bibfnamefont
			{D.}~\bibnamefont {Hu}}, \bibinfo {author} {\bibfnamefont {X.}~\bibnamefont
			{Peng}},\ and\ \bibinfo {author} {\bibfnamefont {D.}~\bibnamefont {Budker}},\
	}\bibfield  {title} {\bibinfo {title} {Search for exotic spin-dependent
			interactions with a spin-based amplifier},\ }\href
	{https://doi.org/10.1126/sciadv.abi9535} {\bibfield  {journal} {\bibinfo
			{journal} {Sci. Adv.}\ }\textbf {\bibinfo {volume} {7}},\ \bibinfo {pages}
		{eabi9535} (\bibinfo {year} {2021})}\BibitemShut {NoStop}%
	\bibitem [{\citenamefont {Stadnik}\ \emph {et~al.}(2018)\citenamefont
		{Stadnik}, \citenamefont {Dzuba},\ and\ \citenamefont
		{Flambaum}}]{stadnik2018improved}%
	\BibitemOpen
	\bibfield  {author} {\bibinfo {author} {\bibfnamefont {Y.~V.}\ \bibnamefont
			{Stadnik}}, \bibinfo {author} {\bibfnamefont {V.~A.}\ \bibnamefont {Dzuba}},\
		and\ \bibinfo {author} {\bibfnamefont {V.~V.}\ \bibnamefont {Flambaum}},\
	}\bibfield  {title} {\bibinfo {title} {Improved {{Limits}} on
			{{Axionlike-Particle-Mediated}} ${{P}}$, ${{T}}$-{{Violating Interactions}}
			between {{Electrons}} and {{Nucleons}} from {{Electric Dipole Moments}} of
			{{Atoms}} and {{Molecules}}},\ }\href
	{https://doi.org/10.1103/PhysRevLett.120.013202} {\bibfield  {journal}
		{\bibinfo  {journal} {Phys. Rev. Lett.}\ }\textbf {\bibinfo {volume} {120}},\
		\bibinfo {pages} {013202} (\bibinfo {year} {2018})}\BibitemShut {NoStop}%
	\bibitem [{\citenamefont {Maze}\ \emph {et~al.}(2008)\citenamefont {Maze},
		\citenamefont {Stanwix}, \citenamefont {Hodges}, \citenamefont {Hong},
		\citenamefont {Taylor}, \citenamefont {Cappellaro}, \citenamefont {Jiang},
		\citenamefont {Dutt}, \citenamefont {Togan}, \citenamefont {Zibrov},
		\citenamefont {Yacoby}, \citenamefont {Walsworth},\ and\ \citenamefont
		{Lukin}}]{maze2008nanoscale}%
	\BibitemOpen
	\bibfield  {author} {\bibinfo {author} {\bibfnamefont {J.~R.}\ \bibnamefont
			{Maze}}, \bibinfo {author} {\bibfnamefont {P.~L.}\ \bibnamefont {Stanwix}},
		\bibinfo {author} {\bibfnamefont {J.~S.}\ \bibnamefont {Hodges}}, \bibinfo
		{author} {\bibfnamefont {S.}~\bibnamefont {Hong}}, \bibinfo {author}
		{\bibfnamefont {J.~M.}\ \bibnamefont {Taylor}}, \bibinfo {author}
		{\bibfnamefont {P.}~\bibnamefont {Cappellaro}}, \bibinfo {author}
		{\bibfnamefont {L.}~\bibnamefont {Jiang}}, \bibinfo {author} {\bibfnamefont
			{M.~V.~G.}\ \bibnamefont {Dutt}}, \bibinfo {author} {\bibfnamefont
			{E.}~\bibnamefont {Togan}}, \bibinfo {author} {\bibfnamefont {A.~S.}\
			\bibnamefont {Zibrov}}, \bibinfo {author} {\bibfnamefont {A.}~\bibnamefont
			{Yacoby}}, \bibinfo {author} {\bibfnamefont {R.~L.}\ \bibnamefont
			{Walsworth}},\ and\ \bibinfo {author} {\bibfnamefont {M.~D.}\ \bibnamefont
			{Lukin}},\ }\bibfield  {title} {\bibinfo {title} {Nanoscale magnetic sensing
			with an individual electronic spin in diamond},\ }\href
	{https://doi.org/10.1038/nature07279} {\bibfield  {journal} {\bibinfo
			{journal} {Nature}\ }\textbf {\bibinfo {volume} {455}},\ \bibinfo {pages}
		{644} (\bibinfo {year} {2008})}\BibitemShut {NoStop}%
	\bibitem [{\citenamefont {Degen}\ \emph {et~al.}(2017)\citenamefont {Degen},
		\citenamefont {Reinhard},\ and\ \citenamefont
		{Cappellaro}}]{degen2017quantum}%
	\BibitemOpen
	\bibfield  {author} {\bibinfo {author} {\bibfnamefont {C.~L.}\ \bibnamefont
			{Degen}}, \bibinfo {author} {\bibfnamefont {F.}~\bibnamefont {Reinhard}},\
		and\ \bibinfo {author} {\bibfnamefont {P.}~\bibnamefont {Cappellaro}},\
	}\bibfield  {title} {\bibinfo {title} {Quantum sensing},\ }\href
	{https://doi.org/10.1103/RevModPhys.89.035002} {\bibfield  {journal}
		{\bibinfo  {journal} {Rev. Mod. Phys.}\ }\textbf {\bibinfo {volume} {89}},\
		\bibinfo {pages} {035002} (\bibinfo {year} {2017})}\BibitemShut {NoStop}%
	\bibitem [{\citenamefont {Liang}\ \emph {et~al.}(2023)\citenamefont {Liang},
		\citenamefont {Jiao}, \citenamefont {Huang}, \citenamefont {Yu},
		\citenamefont {Ye}, \citenamefont {Wang}, \citenamefont {Xie}, \citenamefont
		{Cai}, \citenamefont {Rong},\ and\ \citenamefont {Du}}]{liang2023new}%
	\BibitemOpen
	\bibfield  {author} {\bibinfo {author} {\bibfnamefont {H.}~\bibnamefont
			{Liang}}, \bibinfo {author} {\bibfnamefont {M.}~\bibnamefont {Jiao}},
		\bibinfo {author} {\bibfnamefont {Y.}~\bibnamefont {Huang}}, \bibinfo
		{author} {\bibfnamefont {P.}~\bibnamefont {Yu}}, \bibinfo {author}
		{\bibfnamefont {X.}~\bibnamefont {Ye}}, \bibinfo {author} {\bibfnamefont
			{Y.}~\bibnamefont {Wang}}, \bibinfo {author} {\bibfnamefont {Y.}~\bibnamefont
			{Xie}}, \bibinfo {author} {\bibfnamefont {Y.-F.}\ \bibnamefont {Cai}},
		\bibinfo {author} {\bibfnamefont {X.}~\bibnamefont {Rong}},\ and\ \bibinfo
		{author} {\bibfnamefont {J.}~\bibnamefont {Du}},\ }\bibfield  {title}
	{\bibinfo {title} {New constraints on exotic spin-dependent interactions with
			an ensemble-{{NV-diamond}} magnetometer},\ }\href
	{https://doi.org/10.1093/nsr/nwac262} {\bibfield  {journal} {\bibinfo
			{journal} {Natl. Sci. Rev.}\ }\textbf {\bibinfo {volume} {10}},\ \bibinfo
		{pages} {nwac262} (\bibinfo {year} {2023})}\BibitemShut {NoStop}%
	\bibitem [{\citenamefont {Wu}\ \emph {et~al.}(2023)\citenamefont {Wu},
		\citenamefont {Liang}, \citenamefont {Jiao}, \citenamefont {Cai},
		\citenamefont {Duan}, \citenamefont {Wang}, \citenamefont {Rong},\ and\
		\citenamefont {Du}}]{wu2023improved}%
	\BibitemOpen
	\bibfield  {author} {\bibinfo {author} {\bibfnamefont {D.}~\bibnamefont
			{Wu}}, \bibinfo {author} {\bibfnamefont {H.}~\bibnamefont {Liang}}, \bibinfo
		{author} {\bibfnamefont {M.}~\bibnamefont {Jiao}}, \bibinfo {author}
		{\bibfnamefont {Y.-F.}\ \bibnamefont {Cai}}, \bibinfo {author} {\bibfnamefont
			{C.-K.}\ \bibnamefont {Duan}}, \bibinfo {author} {\bibfnamefont
			{Y.}~\bibnamefont {Wang}}, \bibinfo {author} {\bibfnamefont {X.}~\bibnamefont
			{Rong}},\ and\ \bibinfo {author} {\bibfnamefont {J.}~\bibnamefont {Du}},\
	}\bibfield  {title} {\bibinfo {title} {Improved {{Limits}} on an {{Exotic
					Spin-}} and {{Velocity-Dependent Interaction}} at the {{Micrometer Scale}}
			with an {{Ensemble-NV-Diamond Magnetometer}}},\ }\href
	{https://doi.org/10.1103/PhysRevLett.131.071801} {\bibfield  {journal}
		{\bibinfo  {journal} {Phys. Rev. Lett.}\ }\textbf {\bibinfo {volume} {131}},\
		\bibinfo {pages} {071801} (\bibinfo {year} {2023})}\BibitemShut {NoStop}%
	\bibitem [{\citenamefont {Fayet}(1986)}]{fayet1986fifth}%
	\BibitemOpen
	\bibfield  {author} {\bibinfo {author} {\bibfnamefont {P.}~\bibnamefont
			{Fayet}},\ }\bibfield  {title} {\bibinfo {title} {The fifth interaction in
			grand-unified theories: {{A}} new force acting mostly on neutrons and
			particle spins},\ }\href {https://doi.org/10.1016/0370-2693(86)90271-6}
	{\bibfield  {journal} {\bibinfo  {journal} {Phys. Lett. B}\ }\textbf
		{\bibinfo {volume} {172}},\ \bibinfo {pages} {363} (\bibinfo {year}
		{1986})}\BibitemShut {NoStop}%
	\bibitem [{\citenamefont {Schirhagl}\ \emph {et~al.}(2014)\citenamefont
		{Schirhagl}, \citenamefont {Chang}, \citenamefont {Loretz},\ and\
		\citenamefont {Degen}}]{schirhagl2014nitrogenvacancy}%
	\BibitemOpen
	\bibfield  {author} {\bibinfo {author} {\bibfnamefont {R.}~\bibnamefont
			{Schirhagl}}, \bibinfo {author} {\bibfnamefont {K.}~\bibnamefont {Chang}},
		\bibinfo {author} {\bibfnamefont {M.}~\bibnamefont {Loretz}},\ and\ \bibinfo
		{author} {\bibfnamefont {C.~L.}\ \bibnamefont {Degen}},\ }\bibfield  {title}
	{\bibinfo {title} {Nitrogen-{{Vacancy Centers}} in {{Diamond}}: {{Nanoscale
					Sensors}} for {{Physics}} and {{Biology}}},\ }\href
	{https://doi.org/10.1146/annurev-physchem-040513-103659} {\bibfield
		{journal} {\bibinfo  {journal} {Annu. Rev. Phys. Chem.}\ }\textbf {\bibinfo
			{volume} {65}},\ \bibinfo {pages} {83} (\bibinfo {year} {2014})}\BibitemShut
	{NoStop}%
	\bibitem [{\citenamefont {Gruber}\ \emph {et~al.}(1997)\citenamefont {Gruber},
		\citenamefont {Dr{\"a}benstedt}, \citenamefont {Tietz}, \citenamefont
		{Fleury}, \citenamefont {Wrachtrup},\ and\ \citenamefont {von
			Borczyskowski}}]{gruber1997scanning}%
	\BibitemOpen
	\bibfield  {author} {\bibinfo {author} {\bibfnamefont {A.}~\bibnamefont
			{Gruber}}, \bibinfo {author} {\bibfnamefont {A.}~\bibnamefont
			{Dr{\"a}benstedt}}, \bibinfo {author} {\bibfnamefont {C.}~\bibnamefont
			{Tietz}}, \bibinfo {author} {\bibfnamefont {L.}~\bibnamefont {Fleury}},
		\bibinfo {author} {\bibfnamefont {J.}~\bibnamefont {Wrachtrup}},\ and\
		\bibinfo {author} {\bibfnamefont {C.}~\bibnamefont {von Borczyskowski}},\
	}\bibfield  {title} {\bibinfo {title} {Scanning {{Confocal Optical
					Microscopy}} and {{Magnetic Resonance}} on {{Single Defect Centers}}},\
	}\href {https://doi.org/10.1126/science.276.5321.2012} {\bibfield  {journal}
		{\bibinfo  {journal} {Science}\ }\textbf {\bibinfo {volume} {276}},\ \bibinfo
		{pages} {2012} (\bibinfo {year} {1997})}\BibitemShut {NoStop}%
	\bibitem [{\citenamefont {Jelezko}\ and\ \citenamefont
		{Wrachtrup}(2006)}]{jelezko2006single}%
	\BibitemOpen
	\bibfield  {author} {\bibinfo {author} {\bibfnamefont {F.}~\bibnamefont
			{Jelezko}}\ and\ \bibinfo {author} {\bibfnamefont {J.}~\bibnamefont
			{Wrachtrup}},\ }\bibfield  {title} {\bibinfo {title} {Single defect centres
			in diamond: {{A}} review},\ }\href {https://doi.org/10.1002/pssa.200671403}
	{\bibfield  {journal} {\bibinfo  {journal} {Phys. Status Solidi A}\ }\textbf
		{\bibinfo {volume} {203}},\ \bibinfo {pages} {3207} (\bibinfo {year}
		{2006})}\BibitemShut {NoStop}%
	\bibitem [{\citenamefont {Doherty}\ \emph {et~al.}(2012)\citenamefont
		{Doherty}, \citenamefont {Dolde}, \citenamefont {Fedder}, \citenamefont
		{Jelezko}, \citenamefont {Wrachtrup}, \citenamefont {Manson},\ and\
		\citenamefont {Hollenberg}}]{doherty2012theorya}%
	\BibitemOpen
	\bibfield  {author} {\bibinfo {author} {\bibfnamefont {M.~W.}\ \bibnamefont
			{Doherty}}, \bibinfo {author} {\bibfnamefont {F.}~\bibnamefont {Dolde}},
		\bibinfo {author} {\bibfnamefont {H.}~\bibnamefont {Fedder}}, \bibinfo
		{author} {\bibfnamefont {F.}~\bibnamefont {Jelezko}}, \bibinfo {author}
		{\bibfnamefont {J.}~\bibnamefont {Wrachtrup}}, \bibinfo {author}
		{\bibfnamefont {N.~B.}\ \bibnamefont {Manson}},\ and\ \bibinfo {author}
		{\bibfnamefont {L.~C.~L.}\ \bibnamefont {Hollenberg}},\ }\bibfield  {title}
	{\bibinfo {title} {Theory of the ground-state spin of the {$\mathrm{NV}^{-}$}
			center in diamond},\ }\href {https://doi.org/10.1103/PhysRevB.85.205203}
	{\bibfield  {journal} {\bibinfo  {journal} {Phys. Rev. B}\ }\textbf {\bibinfo
			{volume} {85}},\ \bibinfo {pages} {205203} (\bibinfo {year}
		{2012})}\BibitemShut {NoStop}%
	\bibitem [{\citenamefont {Gali}(2019)}]{gali2019initio}%
	\BibitemOpen
	\bibfield  {author} {\bibinfo {author} {\bibfnamefont {{\'A}.}~\bibnamefont
			{Gali}},\ }\bibfield  {title} {\bibinfo {title} {Ab initio theory of the
			nitrogen-vacancy center in diamond},\ }\href
	{https://doi.org/10.1515/nanoph-2019-0154} {\bibfield  {journal} {\bibinfo
			{journal} {Nanophotonics}\ }\textbf {\bibinfo {volume} {8}},\ \bibinfo
		{pages} {1907} (\bibinfo {year} {2019})}\BibitemShut {NoStop}%
	\bibitem [{\citenamefont {Doherty}\ \emph {et~al.}(2013)\citenamefont
		{Doherty}, \citenamefont {Manson}, \citenamefont {Delaney}, \citenamefont
		{Jelezko}, \citenamefont {Wrachtrup},\ and\ \citenamefont
		{Hollenberg}}]{doherty2013nitrogenvacancy}%
	\BibitemOpen
	\bibfield  {author} {\bibinfo {author} {\bibfnamefont {M.~W.}\ \bibnamefont
			{Doherty}}, \bibinfo {author} {\bibfnamefont {N.~B.}\ \bibnamefont {Manson}},
		\bibinfo {author} {\bibfnamefont {P.}~\bibnamefont {Delaney}}, \bibinfo
		{author} {\bibfnamefont {F.}~\bibnamefont {Jelezko}}, \bibinfo {author}
		{\bibfnamefont {J.}~\bibnamefont {Wrachtrup}},\ and\ \bibinfo {author}
		{\bibfnamefont {L.~C.~L.}\ \bibnamefont {Hollenberg}},\ }\bibfield  {title}
	{\bibinfo {title} {The nitrogen-vacancy colour centre in diamond},\ }\href
	{https://doi.org/10.1016/j.physrep.2013.02.001} {\bibfield  {journal}
		{\bibinfo  {journal} {Phys. Rep.}\ }\textbf {\bibinfo {volume} {528}},\
		\bibinfo {pages} {1} (\bibinfo {year} {2013})}\BibitemShut {NoStop}%
	\bibitem [{\citenamefont {Goldman}\ \emph
		{et~al.}(2015{\natexlab{a}})\citenamefont {Goldman}, \citenamefont
		{Sipahigil}, \citenamefont {Doherty}, \citenamefont {Yao}, \citenamefont
		{Bennett}, \citenamefont {Markham}, \citenamefont {Twitchen}, \citenamefont
		{Manson}, \citenamefont {Kubanek},\ and\ \citenamefont
		{Lukin}}]{goldman2015phononinduced}%
	\BibitemOpen
	\bibfield  {author} {\bibinfo {author} {\bibfnamefont {M.~L.}\ \bibnamefont
			{Goldman}}, \bibinfo {author} {\bibfnamefont {A.}~\bibnamefont {Sipahigil}},
		\bibinfo {author} {\bibfnamefont {M.~W.}\ \bibnamefont {Doherty}}, \bibinfo
		{author} {\bibfnamefont {N.~Y.}\ \bibnamefont {Yao}}, \bibinfo {author}
		{\bibfnamefont {S.~D.}\ \bibnamefont {Bennett}}, \bibinfo {author}
		{\bibfnamefont {M.}~\bibnamefont {Markham}}, \bibinfo {author} {\bibfnamefont
			{D.~J.}\ \bibnamefont {Twitchen}}, \bibinfo {author} {\bibfnamefont {N.~B.}\
			\bibnamefont {Manson}}, \bibinfo {author} {\bibfnamefont {A.}~\bibnamefont
			{Kubanek}},\ and\ \bibinfo {author} {\bibfnamefont {M.~D.}\ \bibnamefont
			{Lukin}},\ }\bibfield  {title} {\bibinfo {title} {Phonon-{{Induced Population
					Dynamics}} and {{Intersystem Crossing}} in {{Nitrogen-Vacancy Centers}}},\
	}\href {https://doi.org/10.1103/PhysRevLett.114.145502} {\bibfield  {journal}
		{\bibinfo  {journal} {Phys. Rev. Lett.}\ }\textbf {\bibinfo {volume} {114}},\
		\bibinfo {pages} {145502} (\bibinfo {year} {2015}{\natexlab{a}})}\BibitemShut
	{NoStop}%
	\bibitem [{\citenamefont {Goldman}\ \emph
		{et~al.}(2015{\natexlab{b}})\citenamefont {Goldman}, \citenamefont {Doherty},
		\citenamefont {Sipahigil}, \citenamefont {Yao}, \citenamefont {Bennett},
		\citenamefont {Manson}, \citenamefont {Kubanek},\ and\ \citenamefont
		{Lukin}}]{goldman2015stateselective}%
	\BibitemOpen
	\bibfield  {author} {\bibinfo {author} {\bibfnamefont {M.~L.}\ \bibnamefont
			{Goldman}}, \bibinfo {author} {\bibfnamefont {M.~W.}\ \bibnamefont
			{Doherty}}, \bibinfo {author} {\bibfnamefont {A.}~\bibnamefont {Sipahigil}},
		\bibinfo {author} {\bibfnamefont {N.~Y.}\ \bibnamefont {Yao}}, \bibinfo
		{author} {\bibfnamefont {S.~D.}\ \bibnamefont {Bennett}}, \bibinfo {author}
		{\bibfnamefont {N.~B.}\ \bibnamefont {Manson}}, \bibinfo {author}
		{\bibfnamefont {A.}~\bibnamefont {Kubanek}},\ and\ \bibinfo {author}
		{\bibfnamefont {M.~D.}\ \bibnamefont {Lukin}},\ }\bibfield  {title} {\bibinfo
		{title} {State-selective intersystem crossing in nitrogen-vacancy centers},\
	}\href {https://doi.org/10.1103/PhysRevB.91.165201} {\bibfield  {journal}
		{\bibinfo  {journal} {Phys. Rev. B}\ }\textbf {\bibinfo {volume} {91}},\
		\bibinfo {pages} {165201} (\bibinfo {year} {2015}{\natexlab{b}})}\BibitemShut
	{NoStop}%
	\bibitem [{\citenamefont {Maertz}\ \emph {et~al.}(2010)\citenamefont {Maertz},
		\citenamefont {Wijnheijmer}, \citenamefont {Fuchs}, \citenamefont
		{Nowakowski},\ and\ \citenamefont {Awschalom}}]{maertz2010vector}%
	\BibitemOpen
	\bibfield  {author} {\bibinfo {author} {\bibfnamefont {B.~J.}\ \bibnamefont
			{Maertz}}, \bibinfo {author} {\bibfnamefont {A.~P.}\ \bibnamefont
			{Wijnheijmer}}, \bibinfo {author} {\bibfnamefont {G.~D.}\ \bibnamefont
			{Fuchs}}, \bibinfo {author} {\bibfnamefont {M.~E.}\ \bibnamefont
			{Nowakowski}},\ and\ \bibinfo {author} {\bibfnamefont {D.~D.}\ \bibnamefont
			{Awschalom}},\ }\bibfield  {title} {\bibinfo {title} {Vector magnetic field
			microscopy using nitrogen vacancy centers in diamond},\ }\href
	{https://doi.org/10.1063/1.3337096} {\bibfield  {journal} {\bibinfo
			{journal} {Appl. Phys. Lett.}\ }\textbf {\bibinfo {volume} {96}},\ \bibinfo
		{pages} {092504} (\bibinfo {year} {2010})}\BibitemShut {NoStop}%
	\bibitem [{\citenamefont {Pham}\ \emph {et~al.}(2011)\citenamefont {Pham},
		\citenamefont {Sage}, \citenamefont {Stanwix}, \citenamefont {Yeung},
		\citenamefont {Glenn}, \citenamefont {Trifonov}, \citenamefont {Cappellaro},
		\citenamefont {Hemmer}, \citenamefont {Lukin}, \citenamefont {Park},
		\citenamefont {Yacoby},\ and\ \citenamefont {Walsworth}}]{pham2011magnetica}%
	\BibitemOpen
	\bibfield  {author} {\bibinfo {author} {\bibfnamefont {L.~M.}\ \bibnamefont
			{Pham}}, \bibinfo {author} {\bibfnamefont {D.~L.}\ \bibnamefont {Sage}},
		\bibinfo {author} {\bibfnamefont {P.~L.}\ \bibnamefont {Stanwix}}, \bibinfo
		{author} {\bibfnamefont {T.~K.}\ \bibnamefont {Yeung}}, \bibinfo {author}
		{\bibfnamefont {D.}~\bibnamefont {Glenn}}, \bibinfo {author} {\bibfnamefont
			{A.}~\bibnamefont {Trifonov}}, \bibinfo {author} {\bibfnamefont
			{P.}~\bibnamefont {Cappellaro}}, \bibinfo {author} {\bibfnamefont {P.~R.}\
			\bibnamefont {Hemmer}}, \bibinfo {author} {\bibfnamefont {M.~D.}\
			\bibnamefont {Lukin}}, \bibinfo {author} {\bibfnamefont {H.}~\bibnamefont
			{Park}}, \bibinfo {author} {\bibfnamefont {A.}~\bibnamefont {Yacoby}},\ and\
		\bibinfo {author} {\bibfnamefont {R.~L.}\ \bibnamefont {Walsworth}},\
	}\bibfield  {title} {\bibinfo {title} {Magnetic field imaging with
			nitrogen-vacancy ensembles},\ }\href
	{https://doi.org/10.1088/1367-2630/13/4/045021} {\bibfield  {journal}
		{\bibinfo  {journal} {New J. Phys.}\ }\textbf {\bibinfo {volume} {13}},\
		\bibinfo {pages} {045021} (\bibinfo {year} {2011})}\BibitemShut {NoStop}%
	\bibitem [{\citenamefont {Schloss}\ \emph {et~al.}(2018)\citenamefont
		{Schloss}, \citenamefont {Barry}, \citenamefont {Turner},\ and\ \citenamefont
		{Walsworth}}]{schloss2018simultaneous}%
	\BibitemOpen
	\bibfield  {author} {\bibinfo {author} {\bibfnamefont {J.~M.}\ \bibnamefont
			{Schloss}}, \bibinfo {author} {\bibfnamefont {J.~F.}\ \bibnamefont {Barry}},
		\bibinfo {author} {\bibfnamefont {M.~J.}\ \bibnamefont {Turner}},\ and\
		\bibinfo {author} {\bibfnamefont {R.~L.}\ \bibnamefont {Walsworth}},\
	}\bibfield  {title} {\bibinfo {title} {Simultaneous {{Broadband Vector
					Magnetometry Using Solid-State Spins}}},\ }\href
	{https://doi.org/10.1103/PhysRevApplied.10.034044} {\bibfield  {journal}
		{\bibinfo  {journal} {Phys. Rev. Appl.}\ }\textbf {\bibinfo {volume} {10}},\
		\bibinfo {pages} {034044} (\bibinfo {year} {2018})}\BibitemShut {NoStop}%
	\bibitem [{\citenamefont {Smeltzer}\ \emph {et~al.}(2009)\citenamefont
		{Smeltzer}, \citenamefont {McIntyre},\ and\ \citenamefont
		{Childress}}]{smeltzer2009robust}%
	\BibitemOpen
	\bibfield  {author} {\bibinfo {author} {\bibfnamefont {B.}~\bibnamefont
			{Smeltzer}}, \bibinfo {author} {\bibfnamefont {J.}~\bibnamefont {McIntyre}},\
		and\ \bibinfo {author} {\bibfnamefont {L.}~\bibnamefont {Childress}},\
	}\bibfield  {title} {\bibinfo {title} {Robust control of individual nuclear
			spins in diamond},\ }\href {https://doi.org/10.1103/PhysRevA.80.050302}
	{\bibfield  {journal} {\bibinfo  {journal} {Phys. Rev. A}\ }\textbf {\bibinfo
			{volume} {80}},\ \bibinfo {pages} {050302} (\bibinfo {year}
		{2009})}\BibitemShut {NoStop}%
	\bibitem [{\citenamefont {Pagliero}\ \emph {et~al.}(2014)\citenamefont
		{Pagliero}, \citenamefont {Laraoui}, \citenamefont {Henshaw},\ and\
		\citenamefont {Meriles}}]{pagliero2014recursive}%
	\BibitemOpen
	\bibfield  {author} {\bibinfo {author} {\bibfnamefont {D.}~\bibnamefont
			{Pagliero}}, \bibinfo {author} {\bibfnamefont {A.}~\bibnamefont {Laraoui}},
		\bibinfo {author} {\bibfnamefont {J.~D.}\ \bibnamefont {Henshaw}},\ and\
		\bibinfo {author} {\bibfnamefont {C.~A.}\ \bibnamefont {Meriles}},\
	}\bibfield  {title} {\bibinfo {title} {Recursive polarization of nuclear
			spins in diamond at arbitrary magnetic fields},\ }\href
	{https://doi.org/10.1063/1.4903799} {\bibfield  {journal} {\bibinfo
			{journal} {Appl. Phys. Lett.}\ }\textbf {\bibinfo {volume} {105}},\ \bibinfo
		{pages} {242402} (\bibinfo {year} {2014})}\BibitemShut {NoStop}%
	\bibitem [{\citenamefont {Soshenko}\ \emph {et~al.}(2021)\citenamefont
		{Soshenko}, \citenamefont {Bolshedvorskii}, \citenamefont {Rubinas},
		\citenamefont {Sorokin}, \citenamefont {Smolyaninov}, \citenamefont
		{Vorobyov},\ and\ \citenamefont {Akimov}}]{soshenko2021nuclear}%
	\BibitemOpen
	\bibfield  {author} {\bibinfo {author} {\bibfnamefont {V.~V.}\ \bibnamefont
			{Soshenko}}, \bibinfo {author} {\bibfnamefont {S.~V.}\ \bibnamefont
			{Bolshedvorskii}}, \bibinfo {author} {\bibfnamefont {O.}~\bibnamefont
			{Rubinas}}, \bibinfo {author} {\bibfnamefont {V.~N.}\ \bibnamefont
			{Sorokin}}, \bibinfo {author} {\bibfnamefont {A.~N.}\ \bibnamefont
			{Smolyaninov}}, \bibinfo {author} {\bibfnamefont {V.~V.}\ \bibnamefont
			{Vorobyov}},\ and\ \bibinfo {author} {\bibfnamefont {A.~V.}\ \bibnamefont
			{Akimov}},\ }\bibfield  {title} {\bibinfo {title} {Nuclear {{Spin Gyroscope}}
			based on the {{Nitrogen Vacancy Center}} in {{Diamond}}},\ }\href
	{https://doi.org/10.1103/PhysRevLett.126.197702} {\bibfield  {journal}
		{\bibinfo  {journal} {Phys. Rev. Lett.}\ }\textbf {\bibinfo {volume} {126}},\
		\bibinfo {pages} {197702} (\bibinfo {year} {2021})}\BibitemShut {NoStop}%
	\bibitem [{\citenamefont {Soshenko}\ \emph {et~al.}(2023)\citenamefont
		{Soshenko}, \citenamefont {Cojocaru}, \citenamefont {Bolshedvorskii},
		\citenamefont {Rubinas}, \citenamefont {Sorokin}, \citenamefont
		{Smolyaninov},\ and\ \citenamefont {Akimov}}]{soshenko2023optimal}%
	\BibitemOpen
	\bibfield  {author} {\bibinfo {author} {\bibfnamefont {V.~V.}\ \bibnamefont
			{Soshenko}}, \bibinfo {author} {\bibfnamefont {I.~S.}\ \bibnamefont
			{Cojocaru}}, \bibinfo {author} {\bibfnamefont {S.~V.}\ \bibnamefont
			{Bolshedvorskii}}, \bibinfo {author} {\bibfnamefont {O.~R.}\ \bibnamefont
			{Rubinas}}, \bibinfo {author} {\bibfnamefont {V.~N.}\ \bibnamefont
			{Sorokin}}, \bibinfo {author} {\bibfnamefont {A.~N.}\ \bibnamefont
			{Smolyaninov}},\ and\ \bibinfo {author} {\bibfnamefont {A.~V.}\ \bibnamefont
			{Akimov}},\ }\bibfield  {title} {\bibinfo {title} {Optimal {{Microwave
					Control Pulse}} for {{Nuclear Spin Polarization}} and {{Readout}} in {{Dense
					Nitrogen-Vacancy Ensembles}} in {{Diamond}}},\ }\href
	{https://doi.org/10.1002/pssb.202200613} {\bibfield  {journal} {\bibinfo
			{journal} {Phys. Status Solidi B}\ }\textbf {\bibinfo {volume} {260}},\
		\bibinfo {pages} {2200613} (\bibinfo {year} {2023})}\BibitemShut {NoStop}%
	\bibitem [{\citenamefont {Hahn}(1950)}]{hahn1950spin}%
	\BibitemOpen
	\bibfield  {author} {\bibinfo {author} {\bibfnamefont {E.~L.}\ \bibnamefont
			{Hahn}},\ }\bibfield  {title} {\bibinfo {title} {Spin {{Echoes}}},\ }\href
	{https://doi.org/10.1103/PhysRev.80.580} {\bibfield  {journal} {\bibinfo
			{journal} {Phys. Rev.}\ }\textbf {\bibinfo {volume} {80}},\ \bibinfo {pages}
		{580} (\bibinfo {year} {1950})}\BibitemShut {NoStop}%
	\bibitem [{\citenamefont {Bauch}\ \emph {et~al.}(2020)\citenamefont {Bauch},
		\citenamefont {Singh}, \citenamefont {Lee}, \citenamefont {Hart},
		\citenamefont {Schloss}, \citenamefont {Turner}, \citenamefont {Barry},
		\citenamefont {Pham}, \citenamefont {{Bar-Gill}}, \citenamefont {Yelin},\
		and\ \citenamefont {Walsworth}}]{bauch2020decoherence}%
	\BibitemOpen
	\bibfield  {author} {\bibinfo {author} {\bibfnamefont {E.}~\bibnamefont
			{Bauch}}, \bibinfo {author} {\bibfnamefont {S.}~\bibnamefont {Singh}},
		\bibinfo {author} {\bibfnamefont {J.}~\bibnamefont {Lee}}, \bibinfo {author}
		{\bibfnamefont {C.~A.}\ \bibnamefont {Hart}}, \bibinfo {author}
		{\bibfnamefont {J.~M.}\ \bibnamefont {Schloss}}, \bibinfo {author}
		{\bibfnamefont {M.~J.}\ \bibnamefont {Turner}}, \bibinfo {author}
		{\bibfnamefont {J.~F.}\ \bibnamefont {Barry}}, \bibinfo {author}
		{\bibfnamefont {L.~M.}\ \bibnamefont {Pham}}, \bibinfo {author}
		{\bibfnamefont {N.}~\bibnamefont {{Bar-Gill}}}, \bibinfo {author}
		{\bibfnamefont {S.~F.}\ \bibnamefont {Yelin}},\ and\ \bibinfo {author}
		{\bibfnamefont {R.~L.}\ \bibnamefont {Walsworth}},\ }\bibfield  {title}
	{\bibinfo {title} {Decoherence of ensembles of nitrogen-vacancy centers in
			diamond},\ }\href {https://doi.org/10.1103/PhysRevB.102.134210} {\bibfield
		{journal} {\bibinfo  {journal} {Phys. Rev. B}\ }\textbf {\bibinfo {volume}
			{102}},\ \bibinfo {pages} {134210} (\bibinfo {year} {2020})}\BibitemShut
	{NoStop}%
	\bibitem [{\citenamefont {Engel}\ and\ \citenamefont
		{Vogel}(1989)}]{engel1989spindependent}%
	\BibitemOpen
	\bibfield  {author} {\bibinfo {author} {\bibfnamefont {J.}~\bibnamefont
			{Engel}}\ and\ \bibinfo {author} {\bibfnamefont {P.}~\bibnamefont {Vogel}},\
	}\bibfield  {title} {\bibinfo {title} {Spin-dependent cross sections of
			weakly interacting massive particles on nuclei},\ }\href
	{https://doi.org/10.1103/PhysRevD.40.3132} {\bibfield  {journal} {\bibinfo
			{journal} {Phys. Rev. D}\ }\textbf {\bibinfo {volume} {40}},\ \bibinfo
		{pages} {3132} (\bibinfo {year} {1989})}\BibitemShut {NoStop}%
	\bibitem [{\citenamefont {Kimball}(2015)}]{kimball2015nucleara}%
	\BibitemOpen
	\bibfield  {author} {\bibinfo {author} {\bibfnamefont {D.~F.~J.}\
			\bibnamefont {Kimball}},\ }\bibfield  {title} {\bibinfo {title} {Nuclear spin
			content and constraints on exotic spin-dependent couplings},\ }\href
	{https://doi.org/10.1088/1367-2630/17/7/073008} {\bibfield  {journal}
		{\bibinfo  {journal} {New J. Phys.}\ }\textbf {\bibinfo {volume} {17}},\
		\bibinfo {pages} {073008} (\bibinfo {year} {2015})}\BibitemShut {NoStop}%
	\bibitem [{\citenamefont {Barry}\ \emph {et~al.}(2020)\citenamefont {Barry},
		\citenamefont {Schloss}, \citenamefont {Bauch}, \citenamefont {Turner},
		\citenamefont {Hart}, \citenamefont {Pham},\ and\ \citenamefont
		{Walsworth}}]{barry2020sensitivity}%
	\BibitemOpen
	\bibfield  {author} {\bibinfo {author} {\bibfnamefont {J.~F.}\ \bibnamefont
			{Barry}}, \bibinfo {author} {\bibfnamefont {J.~M.}\ \bibnamefont {Schloss}},
		\bibinfo {author} {\bibfnamefont {E.}~\bibnamefont {Bauch}}, \bibinfo
		{author} {\bibfnamefont {M.~J.}\ \bibnamefont {Turner}}, \bibinfo {author}
		{\bibfnamefont {C.~A.}\ \bibnamefont {Hart}}, \bibinfo {author}
		{\bibfnamefont {L.~M.}\ \bibnamefont {Pham}},\ and\ \bibinfo {author}
		{\bibfnamefont {R.~L.}\ \bibnamefont {Walsworth}},\ }\bibfield  {title}
	{\bibinfo {title} {Sensitivity optimization for {{NV-diamond}}
			magnetometry},\ }\href {https://doi.org/10.1103/RevModPhys.92.015004}
	{\bibfield  {journal} {\bibinfo  {journal} {Rev. Mod. Phys.}\ }\textbf
		{\bibinfo {volume} {92}},\ \bibinfo {pages} {015004} (\bibinfo {year}
		{2020})}\BibitemShut {NoStop}%
	\bibitem [{\citenamefont {Hopper}\ \emph {et~al.}(2018)\citenamefont {Hopper},
		\citenamefont {Shulevitz},\ and\ \citenamefont {Bassett}}]{hopper2018spin}%
	\BibitemOpen
	\bibfield  {author} {\bibinfo {author} {\bibfnamefont {D.~A.}\ \bibnamefont
			{Hopper}}, \bibinfo {author} {\bibfnamefont {H.~J.}\ \bibnamefont
			{Shulevitz}},\ and\ \bibinfo {author} {\bibfnamefont {L.~C.}\ \bibnamefont
			{Bassett}},\ }\bibfield  {title} {\bibinfo {title} {Spin {{Readout
					Techniques}} of the {{Nitrogen-Vacancy Center}} in {{Diamond}}},\ }\href
	{https://doi.org/10.3390/mi9090437} {\bibfield  {journal} {\bibinfo
			{journal} {Micromachines}\ }\textbf {\bibinfo {volume} {9}},\ \bibinfo
		{pages} {437} (\bibinfo {year} {2018})}\BibitemShut {NoStop}%
	\bibitem [{\citenamefont {Wolf}\ \emph {et~al.}(2015)\citenamefont {Wolf},
		\citenamefont {Neumann}, \citenamefont {Nakamura}, \citenamefont {Sumiya},
		\citenamefont {Ohshima}, \citenamefont {Isoya},\ and\ \citenamefont
		{Wrachtrup}}]{wolf2015subpicotesla}%
	\BibitemOpen
	\bibfield  {author} {\bibinfo {author} {\bibfnamefont {T.}~\bibnamefont
			{Wolf}}, \bibinfo {author} {\bibfnamefont {P.}~\bibnamefont {Neumann}},
		\bibinfo {author} {\bibfnamefont {K.}~\bibnamefont {Nakamura}}, \bibinfo
		{author} {\bibfnamefont {H.}~\bibnamefont {Sumiya}}, \bibinfo {author}
		{\bibfnamefont {T.}~\bibnamefont {Ohshima}}, \bibinfo {author} {\bibfnamefont
			{J.}~\bibnamefont {Isoya}},\ and\ \bibinfo {author} {\bibfnamefont
			{J.}~\bibnamefont {Wrachtrup}},\ }\bibfield  {title} {\bibinfo {title}
		{Subpicotesla {{Diamond Magnetometry}}},\ }\href
	{https://doi.org/10.1103/PhysRevX.5.041001} {\bibfield  {journal} {\bibinfo
			{journal} {Phys. Rev. X}\ }\textbf {\bibinfo {volume} {5}},\ \bibinfo {pages}
		{041001} (\bibinfo {year} {2015})}\BibitemShut {NoStop}%
	\bibitem [{\citenamefont {Baryshev}\ and\ \citenamefont
		{Muehle}(2023)}]{baryshev2023scalable}%
	\BibitemOpen
	\bibfield  {author} {\bibinfo {author} {\bibfnamefont {S.~V.}\ \bibnamefont
			{Baryshev}}\ and\ \bibinfo {author} {\bibfnamefont {M.}~\bibnamefont
			{Muehle}},\ }\bibfield  {title} {\bibinfo {title} {Scalable {{Production}}
			and {{Supply}} {{Chain}} of {{Diamond}} {{Wafers}} {{Using}} {{Microwave}}
			{{Plasma}}: {{A}} {{Mini}}-{{Review}}},\ }\href
	{https://doi.org/10.1109/TPS.2023.3339338} {\bibfield  {journal} {\bibinfo
			{journal} {IEEE Transactions on Plasma Science}\ ,\ \bibinfo {pages} {1}}
		(\bibinfo {year} {2023})}\BibitemShut {NoStop}%
	\bibitem [{\citenamefont {Fadeev}\ \emph {et~al.}(2022)\citenamefont {Fadeev},
		\citenamefont {Ficek}, \citenamefont {Kozlov}, \citenamefont {Budker},\ and\
		\citenamefont {Flambaum}}]{fadeev2022pseudovector}%
	\BibitemOpen
	\bibfield  {author} {\bibinfo {author} {\bibfnamefont {P.}~\bibnamefont
			{Fadeev}}, \bibinfo {author} {\bibfnamefont {F.}~\bibnamefont {Ficek}},
		\bibinfo {author} {\bibfnamefont {M.~G.}\ \bibnamefont {Kozlov}}, \bibinfo
		{author} {\bibfnamefont {D.}~\bibnamefont {Budker}},\ and\ \bibinfo {author}
		{\bibfnamefont {V.~V.}\ \bibnamefont {Flambaum}},\ }\bibfield  {title}
	{\bibinfo {title} {Pseudovector and pseudoscalar spin-dependent interactions
			in atoms},\ }\href {https://doi.org/10.1103/PhysRevA.105.022812} {\bibfield
		{journal} {\bibinfo  {journal} {Phys. Rev. A}\ }\textbf {\bibinfo {volume}
			{105}},\ \bibinfo {pages} {022812} (\bibinfo {year} {2022})}\BibitemShut
	{NoStop}%
	\bibitem [{\citenamefont {O’Hare}\ and\ \citenamefont
		{Vitagliano}(2020)}]{ohare2020cornering}%
	\BibitemOpen
	\bibfield  {author} {\bibinfo {author} {\bibfnamefont {C.~A.}\ \bibnamefont
			{O’Hare}}\ and\ \bibinfo {author} {\bibfnamefont {E.}~\bibnamefont
			{Vitagliano}},\ }\bibfield  {title} {\bibinfo {title} {Cornering the axion
			with {{CP}}-violating interactions},\ }\href
	{https://doi.org/10.1103/PhysRevD.102.115026} {\bibfield  {journal} {\bibinfo
			{journal} {Phys. Rev. D}\ }\textbf {\bibinfo {volume} {102}},\ \bibinfo
		{pages} {115026} (\bibinfo {year} {2020})}\BibitemShut {NoStop}%
	\bibitem [{\citenamefont {Arvanitaki}\ and\ \citenamefont
		{Geraci}(2014)}]{arvanitaki2014resonantly}%
	\BibitemOpen
	\bibfield  {author} {\bibinfo {author} {\bibfnamefont {A.}~\bibnamefont
			{Arvanitaki}}\ and\ \bibinfo {author} {\bibfnamefont {A.~A.}\ \bibnamefont
			{Geraci}},\ }\bibfield  {title} {\bibinfo {title} {Resonantly {{Detecting
					Axion-Mediated Forces}} with {{Nuclear Magnetic Resonance}}},\ }\href
	{https://doi.org/10.1103/PhysRevLett.113.161801} {\bibfield  {journal}
		{\bibinfo  {journal} {Phys. Rev. Lett.}\ }\textbf {\bibinfo {volume} {113}},\
		\bibinfo {pages} {161801} (\bibinfo {year} {2014})}\BibitemShut {NoStop}%
\end{thebibliography}

%

\end{document}